\documentclass[prl,aps,showpacs,twocolumn,superscriptaddress]{revtex4}
\usepackage[utf8]{inputenc}
\usepackage[T1]{fontenc}                
\usepackage{amsmath, amsfonts, stmaryrd,dsfont,graphicx,hyperref,bm}
\usepackage{subfigure}
\usepackage[capitalise]{cleveref}
\def\a{\alpha}
\def\b{\beta}
\def\th{\theta}

\def\6{\partial}

\def\p{\pi}

\newcommand{\eq}[1]{\begin{equation}\begin{split}#1\end{split}\end{equation}}

\def\f{\varphi} 
 
\def\l{\lambda}

\begin{document}
\title{Creep of Chiral Domain Walls}
\author{Dion M.F. Hartmann}\email[E-mail adress: ]{d.m.f.hartmann@uu.nl}\affiliation{Institute for Theoretical Physics, Utrecht University, Leuvenlaan 4, NL-3584 CE Utrecht, The Netherlands}
\author {Rembert A. Duine}\affiliation{Institute for Theoretical Physics, Utrecht University, Leuvenlaan 4, NL-3584 CE Utrecht, The Netherlands}\affiliation{Department of Applied Physics, Eindhoven University of Technology, P.O. Box 513, 5600 MB Eindhoven, The Netherlands}
\author{Mariëlle J. Meijer}\affiliation{Department of Applied Physics, Eindhoven University of Technology, P.O. Box 513, 5600 MB Eindhoven, The Netherlands}
\author{Henk J.M. Swagten}\affiliation{Department of Applied Physics, Eindhoven University of Technology, P.O. Box 513, 5600 MB Eindhoven, The Netherlands}
\author{Reinoud Lavrijsen}\affiliation{Department of Applied Physics, Eindhoven University of Technology, P.O. Box 513, 5600 MB Eindhoven, The Netherlands}
\begin{abstract}
Recent experimental studies of magnetic domain expansion under easy-axis drive fields in materials with a perpendicular magnetic anisotropy have shown that the domain wall velocity is asymmetric as a function of an external in plane magnetic field. This is understood as a consequence of the inversion asymmetry of the system, yielding a finite chiral Dzyaloshinskii-Moriya interaction. Numerous attempts have been made to explain these observations using creep theory, but, in doing so, these have not included all contributions to the domain wall energy or have introduced additional free parameters. 
In this article we present a theory for creep motion of chiral domain walls in the creep regime that includes the most important contributions to the domain-wall energy and does not introduce new free parameters beyond the usual parameters that are included in the micromagnetic energy.
Furthermore, we present experimental measurements of domain wall velocities as a function of in-plane field that are well decribed by our model, and from which material properties such as the strength of the Dzyaloshinskii-Moriya interaction and the demagnetization field are extracted. 
\end{abstract}
\maketitle
\textit{Introduction.} ---
The interest in nanomagnetic materials has grown steadily since magnetic storage devices, such as the racetrack memory, were proposed as a new tool to meet the ever increasing demand for computer storage capacity \cite{Par08,Moo15,All05,Hay08}. For such applications the domain wall (DW) chirality is an important parameter as it affects the speed and direction of DW motion. The interfacial Dzyaloshinskii-Moriya-interaction (DMI) \cite{Dzy58,Mor60} arises from perpendicular inversion asymmetry in the system and affects the DW chirality. Hence it is of paramount importance to be able to measure the magnitude of the DMI using a simple experimental method. The interfacial DMI is modeled as an effective field that lies in-plane (IP) and is always perpendicular to the domain wall (DW) normal, hence preferring a Néel wall \cite{Fer80}. Superpositioning the DMI field with an externally applied IP magnetic field could provide means of measuring it. This has lead to a boom of experimental studies on DW dynamics under the influence of an IP magnetic field \cite{Je13,Bou13,Lav15,Jue15,Van15,Lau16,Sok17,Kim17,Sok18,Kim18}. 

There are several regimes of DW dynamics, determined by the strength of the DW driving force compared to the pinning force. In the flow regime the driving force is significantly higher than the pinning force and in this regime IP magnetic fields and DMI is succesfully modeled by means of the Landau-Lifschitz-Gilbert equation \cite{Lan55,Li04,Thi12}. In the creep-regime however, the DW is considered to be mostly pinned and in local equilibrium and has a net displacement because the bias is assisted by thermal fluctuations. 

The creep model was successfully implemented to interpret magnetic domain growth driven by an external magnetic field $H_z$ in the direction of the magnetization of one of the domains, resulting in the famous universal creep law for the DW velocity $v$: $\ln(v)\propto H_z^{-1/4}$ \cite{Lem98}. When introducing a magnetic field perpendicular to the magnetization direction of the domains, a modification to this creep law was proposed: $\ln(v)\propto (E_\textrm{el}/H_z)^{1/4}$, where $E_{el}$ is the elasticity of the DW \cite{Je13}. This modification turned out to described the experimental finding well for small IP magnetic fields, but is not able to describe the high-field region \cite{Lav15}. Recent attempts to improve the theoretical model exposed the dispersive nature of the elasticity but compromised on universality as extra free parameters were introduced that do not occur in the micromagnetic energy functional. One example of such a parameter is the length scale $L$ of the DW segments over which the creep motion occurs.  In previous theories for DW motion in the creep regime this length scale does not enter the prediction for the velocity. It is, however, treated as a free fit parameter in Ref.~[\onlinecite{Sok17}], which makes a direct comparison with experiment hard. Furthermore, chiral damping was proposed to explain the asymmetric component of the velocity profiles \cite{Jue15,Kim16,Kim17,Sok18}. We contend however that in the quasi-static creep regime dynamic effects such as chiral damping should not play a significant role. 

In this paper we construct a theory for motion of chiral domain walls in the creep regime which does not involve the free parameters introduced in Ref.~[\onlinecite{Sok17}]. We use it to interpret our experimental data on the DW velocity as a function of the IP magnetic field. We show that our model allows for quantitative determination of the strength of the interfacial DMI from field-driven DW creep measurements. 


\textit{Model.} ---
In the regime where some elastic manifold is pinned stronger than the applied driving force, one can still obtain a net motion from the combined effects of driving forces, thermal fluctuations and elasticity. Such motion is called creep. The creep model has been used to successfully describe vortex dynamics in superconductors (for a review see  Ref.~[\onlinecite{Bla94rev}]). Based on this work Lemerle \textit{et al.} have shown that a DW in a thin magnetic film with PMA can be modelled successfully within this creep framework \cite{Lem98}.

\begin{figure}[t!]
	\includegraphics[width=\linewidth]{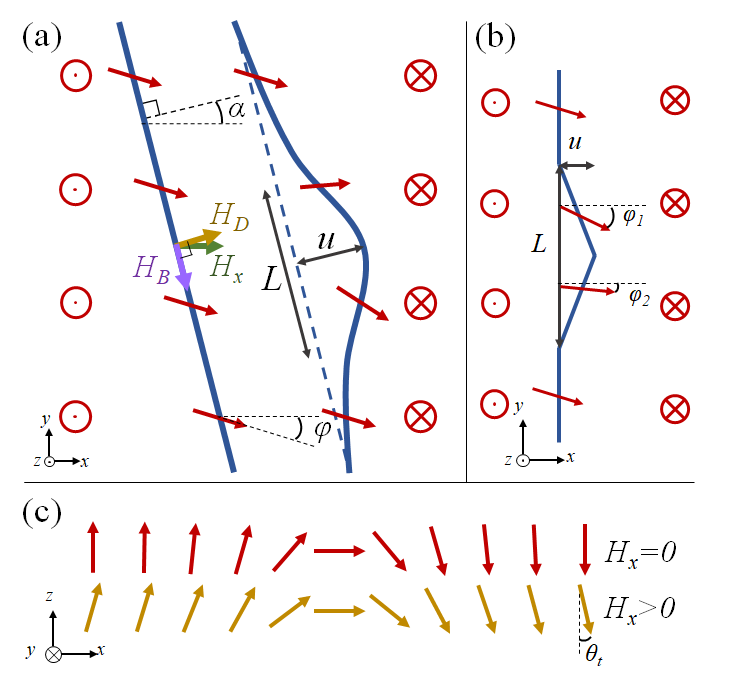}
	\caption{\label{fig:sys} \textbf{(a)}: Top view of a DW (blue lines) that gets deformed over a length $L$  and displaced over a distance $u$ due to a thermal fluctuation. The DW can be tilted over an angle $\a$. The magnetization is indicated by the red vectors, which at the DW are characterized by the IP angle $\f$. Note that the IP magnetization changes due to the displacement, affecting the elasticity. The IP magnetic field $H_x$ (green) as well as the effective DMI field $H_\textrm{D}$ (yellow) and effective Bloch field $H_\textrm{B}$ (purple) are indicated locally. \textbf{(b)}: Model to describe the deformation. \textbf{(c)}: When an IP magnetic field is applied to a sample with PMA, the magnetization inside a domain tilts towards the IP magnetic field by an angle $\th_t$ determined by the balance of PMA and IP magnetic field $\b=M_\textrm{S} H_x \cos(\f)/K_\textrm{P}=\sin(\th_t)$ (orange) compared to the $\b=0$ case (red).}
\end{figure} 

In \cref{fig:sys} (a) the deformation of a DW due to a thermal fluctuation in the presence of an easy axis driving field $H_z$ is illustrated. If the size of the deformation $L$ is relatively small, the gained Zeeman energy from the driving field will be small relative to the elastic energy cost and the deformation cannot grow. But for increasing $L$ the Zeeman energy starts to dominate and deformations can grow. The deformations can be seen as nucleations whose chance of survival is determined by their size. For such a nucleation process, Arrhenius' law tells us that the rate at which these surviving deformations will occur is determined by the height of the energy barrier $F_\textrm{b}$ (i.e. the free energy at the tipping point): $\ln(v)\propto -F_\textrm{b}/(k_\textrm{B} T)$ \cite{Arr89a,Arr89b}. 

Thus, to model the DW dynamics we need to determine the free energy $F(L)$ of the DW segment as a function of $L$ and optimize it to find $F_\textrm{b}=\max_{L}F(L)$. The free energy is composed of the elastic energy cost and the Zeeman energy gain, which depend not only on $L$, but also on the DW displacement $u$: $F(u,L)=E_\textrm{el}(u,L)+E_\textrm{Zeeman}(u,L)$. To express $u$ in terms of $L$ we use $u(L)=u_\textrm{c}(L/L_\textrm{c})^{2/3}$ \cite{Lem98,Kar85,Fis91rev}, where $L_\textrm{c}$ is the Larkin length scale determined by minimizing the sum of the elastic and pinning energy density for $u=\xi$, and $u_\textrm{c}$ is a  proportionality constant. Hence, the next step is to determine the elastic energy to be able to compute $L_\textrm{c}$ and express $u$, and thereby $F$, in terms of $L$.


The elastic energy is defined as the difference in internal, i.e. excluding pinning and driving, energy between the domain wall before and after the deformation. Naively this would just be given by the DW energy density times the added length due to the deformation, but due to the application of the external IP magnetic field the DW energy density itself depends on the orientation of the DW with respect to this applied field. Furthermore, the IP magnetization of the sample at the DW is affected by the exchange interaction. This induces an extra energy cost of bending the DW.

With these considerations in mind, only the elongation of the DW due to the deformation does not provide enough information; we need to know the shape of the deformation. Following Blatter \textit{et al.} we model the deformation as an angular shape for simplicity, see \cref{fig:sys} (b) \cite{Bla94rev}. Note that Pellegren \textit{et al.} chose an arc shape \cite{Sok17}, but did not implement the exchange energy cost due to the kink in the connection with the straight DW segments, resulting in unphysical divergences (as demonstrated in the Supplemental Material \cite{SM}) that do not occur in our theory.

As a first approximation we have chosen the IP magnetization of each separate segment to be constant and implement the bending energy cost as a nearest neighbor exchange interaction at the bending points. The energy of the system is then minimized (numerically) over the IP magnetization angle of the two segments.

%
We compute the energy density of the domain wall by inserting the domain-wall solution into the micromagnetic energy functional. For more details see the Supplemental Material \cite{SM}. This yields for the energy per unit length and layer thickness of a straight domain wall: 
	\eq{
		\label{eq:enden}
		\mathcal{E}(\a,\f)=&2\sqrt{1-\b^2}\frac{J}{\l}+M_\textrm{S} \p \l \left(g(\b)H_\textrm{B}\cos^2(\f-\a)\right. \\
		&\left.-f(\b)\left(H_x\cos(\f)+H_\textrm{D}\cos(\f-\a)\right)\right).
	}
The first term is the exchange interaction $J$ over $\l$, the DW thickness. The second term is the demagnetization energy, expressed in terms of the effective Bloch field $H_\textrm{B}$ (this energy favors a Bloch DW, hence the nomenclature), the angle $\a$ between the DW normal and the $x$-axis and the angle $\f$ the IP magnetization at the DW with the direction of the IP magnetic field, see \cref{fig:sys} (a). The third term is the Zeeman energy due to the applied IP magnetic field $H_x$ and the fourth is the DMI expressed in terms of an effective field $H_\textrm{D}$ favoring a Néel type DW. 
The prefactors involving $\b$ incorporate the tilting in the $x$-direction of the magnetic domains due to the external IP magnetic field (see \cref{fig:sys} (c)). The functions $f$ and $g$ are given in the Supplemental Material \cite{SM}.

Similarly, we obtain the Zeeman energy from the driving field $H_z$
\eq{
\label{eq:Eze}
E_\textrm{Zeeman}(u,L)=M_\textrm{S} H_z t u L \sqrt{1-\b^2}.
}
Again, the factor $\sqrt{1-\b}$ comes from the tilted domains as illustrated in \cref{fig:sys} (c). By dividing out $D$ in \cref{eq:enden,eq:Eze}, the relevant dimensionless parameters become $\tilde{J}\equiv J \lambda^{-1} D^{-1}$, $\tilde{H_\textrm{B}}\equiv2H_B/H_D$, $\tilde{H_x}\equiv H_x/H_\textrm{D}$ and $\tilde{H_z}\equiv H_z/H_\textrm{D}$. 

\begin{figure}[t]
	\includegraphics[width=\linewidth]{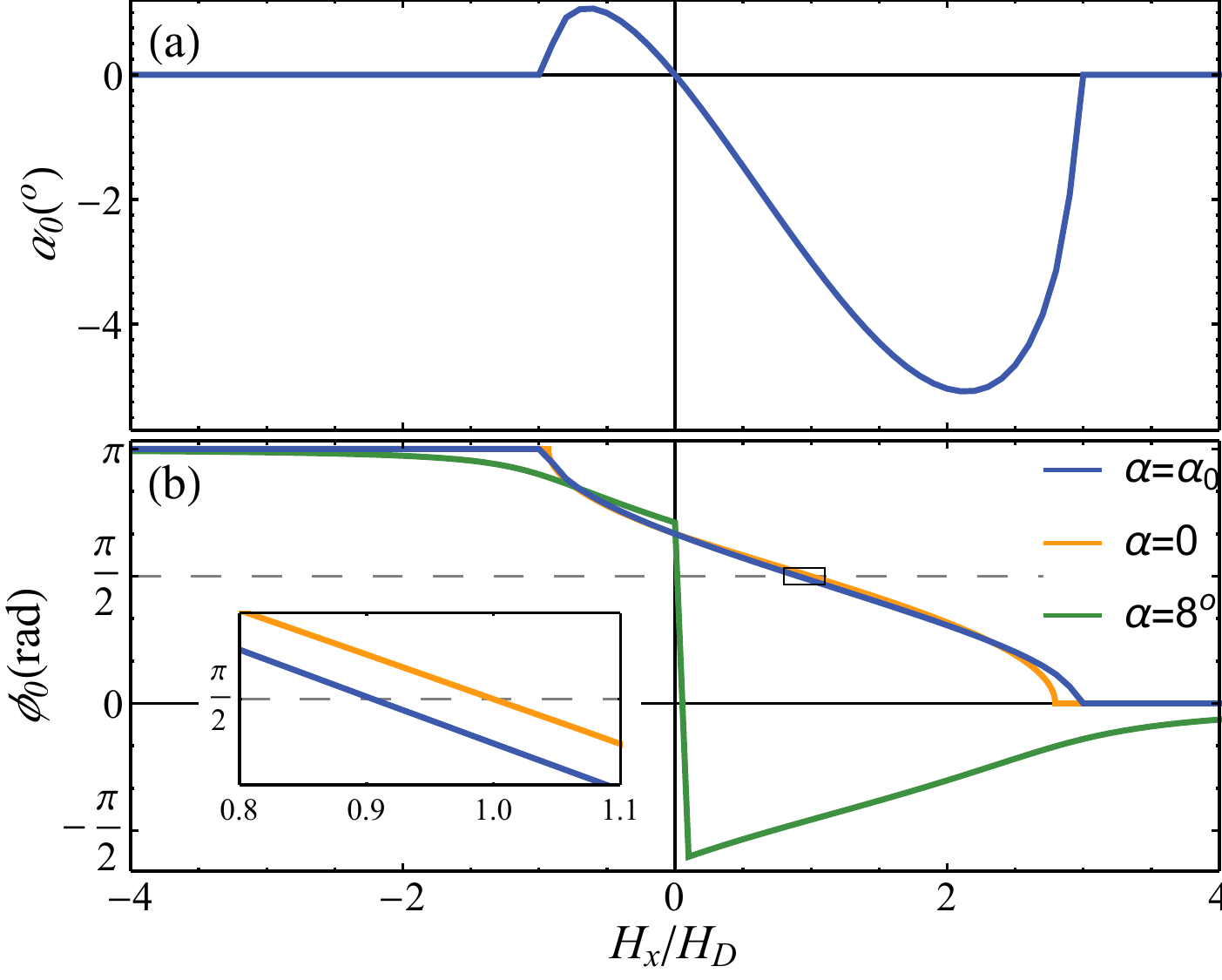}
	\caption{\label{fig:af0} $\a_0$  as a function of the applied IP magnetic field \textbf(a) and the corresponding minimized azimuthal angle of the internal magnetization $\f_0$ \textbf(b). The green curve shows the solution for $\f_0$ when $\a$ is fixed at $8^\circ$, which switches sign at $H_x=0$. The corresponding energy density however, remains continuous and smooth. Note that the green curve does not saturate in but converges to the Néel wall.}
\end{figure}

Using \cref{eq:enden}, we compute the optimal orientation angle of the undeformed DW $\a_0$ and the corresponding internal magnetization IP angle $\f_0$ by minimizing $\mathcal{E}(\a,\f)/\cos(\a)$.
The factor $1/\cos(\a)$ arises because we allow the DW to orient itself with respect to the IP magnetic field at the cost of elongating. For example, a mixed Bloch-Néel DW tilts its normal to better align with the external IP magnetic field. This tilting however would induce a stretching factor of $1/\cos{\a}$, increasing the energy cost. This effect is illustrated in \cref{fig:sys} (a) and the optimal angle $\a_0$ and corresponding minimized angle $\f_0$ are shown as a function of $H_x$ in \cref{fig:af0}. The energy of the unperturbed DW is then given by $Lt\mathcal{E}(\a_0,\f_0)$.

The profile of $\f_0$ shown in \cref{fig:af0} (b) exhibits sharp kinks for both $\a(H_x)=\a_0(H_x)$ and $\a(H_x)=0$. This feature arises because in the energy density of \cref{eq:enden} we neglected higher order anisotropy terms proportional to $\cos^n(\f-\a)$ for $n>2$ which are allowed by symmetry. As a consequence, this simplified energy density yields a sharp transition in DW type from mixed Bloch-Néel to pure Néel at $f(\b)(\tilde{H_x}-1)=g(\b)\tilde{H_\textrm{B}}$ as demonstrated in \cref{fig:af0} where $\f_0$ saturates to $0$ or $\pi$. To effectively include for the higher order terms in the energy density, we adjust the value of $\a$ to some fixed value, e.g. $\a=8^\circ$ as done by Pellegren \textit{et al.}\cite{Sok17}. With this modification, $\f$ is smooth around $\f=0$ or $\f=\pi$ as demonstrated by the green curve in \cref{fig:af0} (b).

To compute the energy of the deformed DW we need to account for a bending energy cost due to exchange interaction. A kink between two DW segments, as illustrated in \cref{fig:sys} (b), gives an energy cost 
\eq{
	\mathcal{E}_\textrm{ben}(\f_1,\f_2)=
	\frac{J\l}{a}(1-\cos(\f_1-\f_2)),
}
with $\f_1$ and $\f_2$ the IP angles of the internal magnetization of the segments. Here, $a$ is the distance between neighboring atoms in the magnetic layer. Due to variations in the lattice structure and to account for non-nearest neighbor interactions, an effective value of $a\sim 1$ nm is used. The effect of $a$ on the DW dynamics is investigated in the Supplemental Material \cite{SM}.

The elastic energy is computed by minimizing over $\f_1$ and $\f_2$:
\begin{widetext}
	\eq{
		\label{eq:eel}
		\frac{E_\textrm{el}(u,L)}{t}=&\min_{\f_1,\f_2}\left[
		\frac{L}{2}\sqrt{1+\left(\frac{2u}{L}\right)^2}\left(
		\mathcal{E}(\a_0+\arctan(2u/L),\f_1)+\mathcal{E}(\a_0-\arctan(2u/L),\f_2)
		\right)
		\right.\\
		&\left.+\frac{J\l}{a}(3-\cos(\f_0-\f_1)-\cos(\f_0-\f_2)-\cos(\f_1-\f_2))
		\right]-L\mathcal{E}(\a_0,\f_0).
	}
\end{widetext}
The first term is the length of each of the two segments of the deformed DW multiplied by their respective energy densities. The second term is the bending energy for the three corners, see \cref{fig:sys} (b). The third term is the energy of the unperturbed DW.

With this expression we compute $L_\textrm{c}$, express $u$ in terms of $L$ and thereby obtain $F(L)=F(u(L),L)$ from which the DW velocity is found as $\ln(v)\propto -F_\textrm{b}/(k_\textrm{B} T)$. For more detail, see the Supplemental Material \cite{SM}. In summary, the derivation of the DW velocity involves multiple optimization steps to determine $\a_0$, $\f_0$, $\f_1$, $\f_2$, $L_\textrm{c}$ and finally $F_\textrm{b}$. Due to the complexity of the elastic energy, these cannot be made analytically. Approximating the elasticity to be proportional to $u^2/L$ does allow for analytic solutions, but these are not able to fully explain recent experimental observations. For example, Je \textit{et al.} approximated \cref{eq:eel} by setting $\a_0=0$, $\f_0=\f_1=\f_2$ and neglecting the $\pm\arctan(2u/L)$ in the first two terms \cite{Je13}. Because Pellegren \textit{et al.} have chosen a different DW profile, we cannot directly compare the expression in \cref{eq:enden} with their results \cite{Sok17}. They do, however, treat $L$ as a free parameter and do not find it by optimization. Moreover, they do not account for the bending costs that we model by the terms involving $\f_0-\f_1$ and $\f_0-\f_1$ in \cref{eq:enden}. The results discussed in the next section are obtained numerically without making further approximations.

\begin{figure}[t!]
	\includegraphics[width=\linewidth]{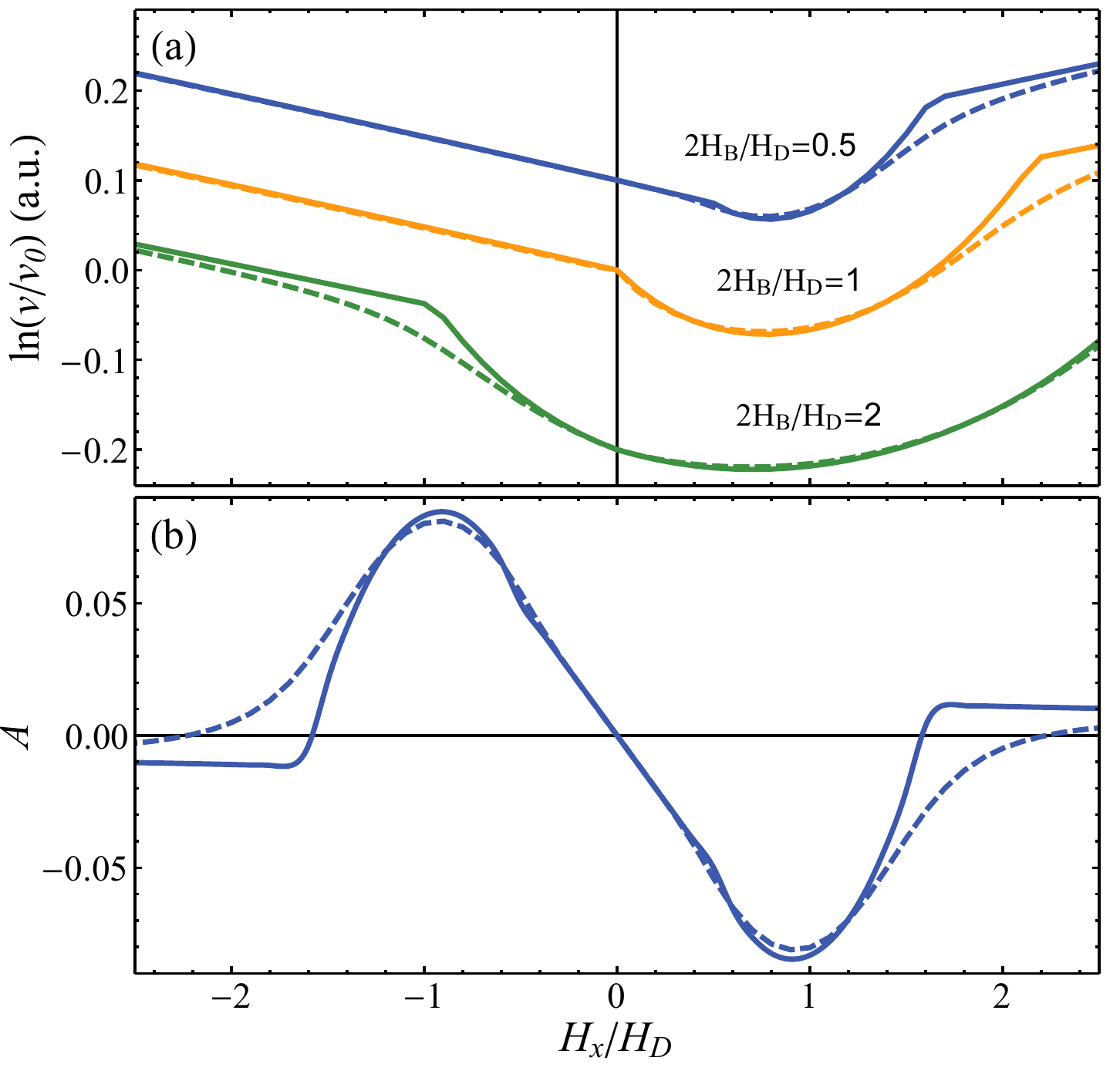}
	\caption{\label{fig:re3} Dependence of the IP magnetic field $\tilde{H_x}$ of the DW velocity \textbf(a) and the assymetric component of the velocity $A$ for $\tilde{H_\textrm{B}}=0.5$ \textbf(b). The profiles in \textbf(a) are given a vertical offset for clarity. The dashed lines represent the result for fixing $\alpha=8^\circ$.}
\end{figure}

\textit{Results.} ---
In \cref{fig:re3}(a) the modeled DW velocity as a function of the applied IP magnetic field $H_x$ is shown for different values of $\tilde{H_\textrm{B}}$ (a). \cref{fig:re3}(b) shows the asymmetric component $A=\ln(v(\uparrow\downarrow)/v(\downarrow\uparrow))$ for $\tilde{H_\textrm{B}}=0.5$. The kinks in the solid lines at $\tilde{H_x}=1\pm\tilde{H_\textrm{B}}$ mark the saturation of internal DW magnetization angle into a Néel wall perpendicular to the IP magnetic field. These are expected from the form of \cref{eq:enden} where we neglected terms $\mathcal{O}(\cos^4(\f))$. The dashed curves are the result of setting $\a=8^\circ$ fixed to compensate for the simplified energy density.

In the high IP magnetic field regime, i.e. $|\tilde{H_x}|>\tilde{H_\textrm{B}}$, the profile straightens out. In this regime the azimuthal angle of the internal magnetization is saturated to align with the IP magnetic field, yielding a Néel DW. Due to this saturation, the orientation dependence of the elasticity no longer varies with further increasing $|H_x|$. As a result, the elasticity becomes isotropic and the logarithmic increase in velocity is linear with $H_x$ solely due to the gained Zeeman energy. 

For $|\tilde{H_x}|\leq\tilde{H_\textrm{B}}$ the DW is mixed Bloch-Néel and the DW velocity provides a distinguishing feature regarding where the steepest slope of the velocity profile with respect to $\tilde{H_x}=0$ and $\tilde{H_x}=1$ is found: When the steepest slope is attained at $|\tilde{H_x}|>1$, then $\tilde{H_\textrm{B}}<1$ (that is, at $\tilde{H_x}=0$ the DW is purely Néel). Otherwise, there is a steep slope around $H_x=0$ (where the DW now is mixed Bloch-Néel). This distinction thus indicates the strength of demagnetization relative to DMI. 

Note that the demonstrated asymmetry of the profile compares well with experiments \cite{Lav15,Hra14,Jue15,Kim17,Kim18,Sok17,Sok18}. Furthermore, the minimal velocity is not attained at $\tilde{H_x}=1$ as in the model of Je \textit{et al.} \cite{Je13}.

Note moreover that the asymmetric velocity component switches sign as $|\tilde{H_x}|$ increases. This feature has been observed experimentally and explained by chiral damping \cite{Jue15,Kim17,Lau16}. In our model there are no chiral damping effects, showing that this feature need not be an indication for chiral damping.

\begin{figure}[t!]
	\includegraphics[width=\linewidth]{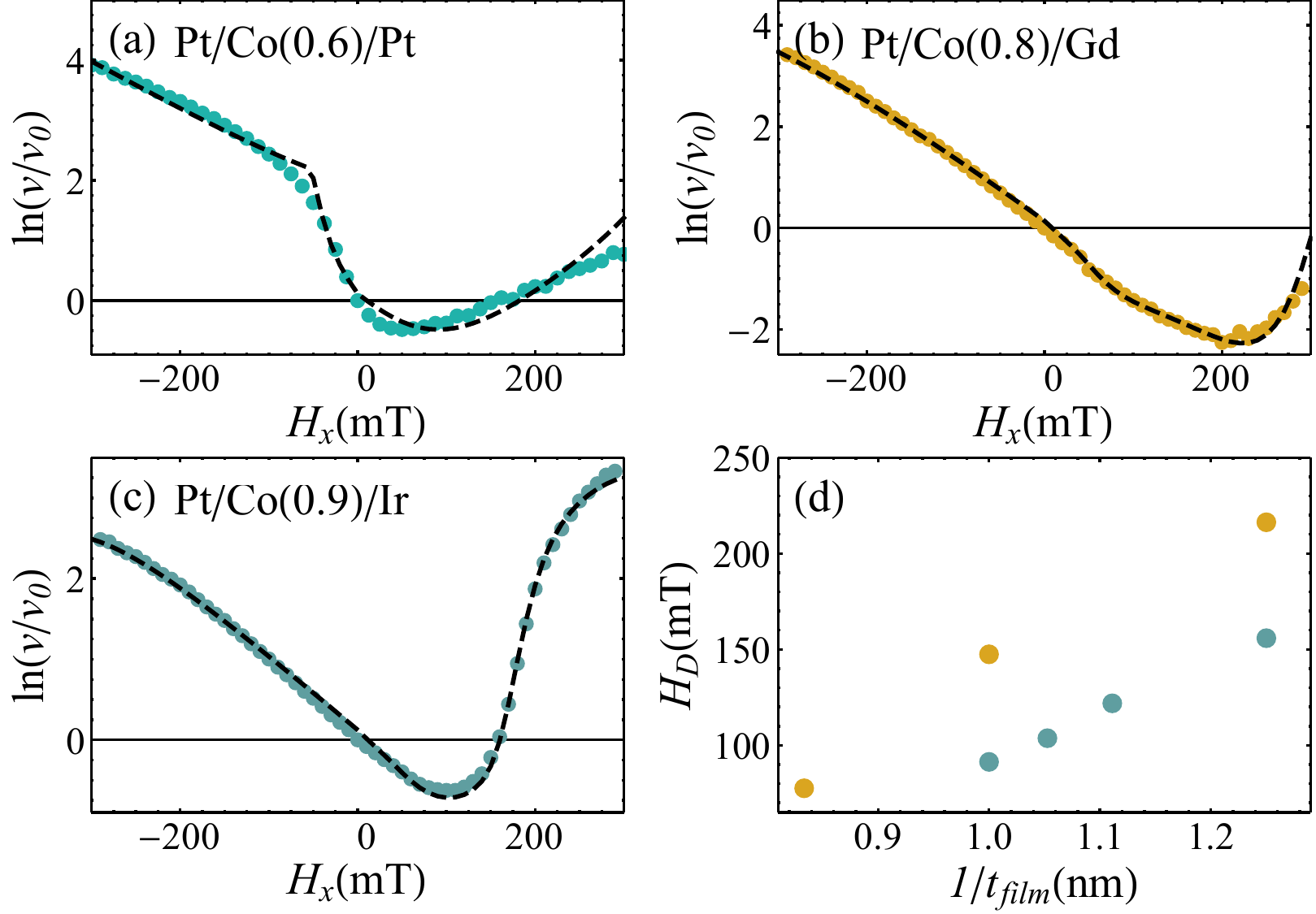}
	\footnotesize
	\begin{tabular}{r|c c c}
		Sample & (a) & (b) & (c) \\
		\hline
		$H_\textrm{D}$(T) & $0.17$ & $0.22$ & $0.12$\\
		$\tilde{H_\textrm{B}}$ & $1.4$ & $0.7$ & $0.6$ \\
		$a_{eff}$(nm) & $1.0$ & $1.7$ & $2.0$
	\end{tabular}
	\caption{\label{fig:fit} Fitted DW velocity curve (dashed line) to experimental data (dots) of three different samples. The data shown in \textbf(b) and \textbf(c) and is obtained for this paper. Details on the samples and measurements as well as the assumed material parameters can be found in the Supplemental Material \cite{SM}. The data in \textbf(a) is from Ref.~[\onlinecite{Lav15}]. The obtained fit parameters are shown in the table. In \textbf(d) the $H_\textrm{D}(t_{film})$ trend for the Pt/Co/Gd (orange) and Pt/Co/Ir (blue) stacks are shown.}
\end{figure}

Finally, we compared and fitted our model to experimental data. The results are shown in \cref{fig:fit} showing good quantitative agreement in a broad variety of samples. We performed measurements on two different samples stacks, see \cref{fig:fit} (b) and (c) (see the Supplemental Material for details on these samples, the method of measurement and more fit results \cite{SM}). Furthermore, we also interpret data from previous research of Ref.~[\onlinecite{Lav15}] in \cref{fig:fit} (a). In \cref{fig:fit} (d) the obtained values for $H_\textrm{D}$ are plotted as a function of the film thickness $t$ and confirm our expectation that $H_\textrm{D}$ should decrease as a function of $t$ \cite{Met07,Thi12,SM}.


\textit{Conclusion.} ---
The DMI and IP magnetic field complexify DW dynamics significantly due to the orientation dependence of elasticity. To grasp and expose this complexity, we defined a model following creep theory and solving the dynamics semi-analytically. The model has a profound sensitivity to DMI and demagnetization. As a result, the model provides a quantitative interpretation of experimental data of DWs that demonstrate asymmetric velocity profiles as a function of $H_x$.

Experimental studies that do not exhibit a (distinctive) kink at $H_x/H_\textrm{D}=\pm 2H_\textrm{B}$ are often fitted with the constant elasticity model proposed by Je \textit{et al.}\cite{Je13} In these studies the measurement range of $H_x$ might not be large enough to expose these kinks. \cref{fig:fit}(b) demonstrates that our model resembles results from the constant elasticity model of Je \textit{et al.} \cite{Je13}, but yields a different value of the DMI

The parameter $\a$ has been set to a fixed value to account for the omission of higher order anisotropy terms in the energy density. As a result the angle $\f$ will not saturate for large $H_x$. Previous research used $\a$ as a fitting parameter as to account for roughness \cite{Sok17}. If roughness forces the DW to tilt, the tilting angle is not fixed to one value. Hence a fixed value of $\a$ should not be interpreted as a physical tilting of the DW.
%

We remark that assuming $\f$ to be constant along an axis normal to the DW is only a first approximation.
For a Mixed Bloch-Néel DW, $\f$ will adjust so that the magnetization aligns with the IP magnetic field well inside the domains, but does not at the DW. 
As $\f$ plays a key role in the DW dynamics, future research could focus on the exact behavior of $\f$.

In recent publications the asymmetric shape of the DW velocity profile as a function of $H_x$ is used as an argument for significant effect of chiral damping on the DW dynamics \cite{Jue15,Kim17,Sok18}. However, our model demonstrates a similar asymmetry without chiral damping. Furthermore, in the quasi-static creep regime dynamic effects such as chiral damping should not affect creep motion.

The comparison experimental data demonstrates the broad applicability of our model. Future research could apply our model to an extensive sample study to investigate the effects of parameters such as layer thickness or growth. 
\begin{acknowledgments}
R.D. is member of the D-ITP consortium, a program of the Dutch Organisation for Scientific Research (NWO) that is funded by the Dutch Ministry of Education, Culture and Science (OCW). This work is funded by the European Research Council
(ERC). This work is part of the research programme of the Foundation for Fundamental Research on Matter (FOM), which is part of the Netherlands Organisation for Scientific Research (NWO).
\end{acknowledgments}
\bibliography{References}
\end{document}


\title{Creep of Chiral Domain Walls\\
		\normalsize Supplemental Material}
	\author{Dion M.F. Hartmann}\email[E-mail adress: ]{d.m.f.hartmann@uu.nl}\affiliation{Institute for Theoretical Physics, Utrecht University, Leuvenlaan 4, NL-3584 CE Utrecht, The Netherlands}
	\author {Rembert A. Duine}\affiliation{Institute for Theoretical Physics, Utrecht University, Leuvenlaan 4, NL-3584 CE Utrecht, The Netherlands}\affiliation{Department of Applied Physics, Eindhoven University of Technology, P.O. Box 513, 5600 MB Eindhoven, The Netherlands}
	\author{Mariëlle J. Meijer}\affiliation{Department of Applied Physics, Eindhoven University of Technology, P.O. Box 513, 5600 MB Eindhoven, The Netherlands}
	\author{Henk J.M. Swagten}\affiliation{Department of Applied Physics, Eindhoven University of Technology, P.O. Box 513, 5600 MB Eindhoven, The Netherlands}
	\author{Reinoud Lavrijsen}\affiliation{Department of Applied Physics, Eindhoven University of Technology, P.O. Box 513, 5600 MB Eindhoven, The Netherlands}
	\begin{abstract}
Recent experimental studies of magnetic domain expansion under easy-axis drive fields in materials with a perpendicular magnetic anisotropy have shown that the domain wall velocity is asymmetric as a function of an external in plane magnetic field. This is understood as a consequence of the inversion asymmetry of the system, yielding a finite chiral Dzyaloshinskii-Moriya interaction. 
In this supplemental material we expand upon the calculations, theories and methods used in the main article. In particular, we elaborate and briefly review creep theory to be self contained.
\end{abstract}
	\pacs{75.60.Ch,75.70.Ak,75.70.Kw,75.78.Fg}
	\maketitle
The reader is referred to the main article for an introduction of the subject and description of the model.
\section{Magnetization profile and energy density}
\label{app:A}
 In this section we derive the micromagnetic energy of a thin system with perpendicular magnetic anisotropy (PMA) and a domain wall (DW), wherewith we compute the spatial magnetization vector profile first for untilted domains and then for tilted domains.  We conclude the section by computing the energy density of a DW with tilted domains.
\subsection{Micromagnetic energy}
Here we sum up the contributions of all considered magnetic effects that make up the energy of a thin ferromagnetic layer with spin inversion asymmetry and PMA. We express each effect in terms of a corresponding energy or (effective) magnetic field. All magnetic fields have units of Tesla and the saturation magnetization is expressed in Ampere per meter (if preferred, one could interchange these two units without further consequences).

Let us denote the magnetization in the ferromagnet by the unit vector in spherical coordinates,
\eq{
	\label{eq:Omega}
	\mathbf{\Omega}=\left(
	\begin{array}{c}
		\cos(\phi)\sin(\theta)\\
		\sin(\phi)\sin(\theta)\\
		\cos(\theta)\\
	\end{array}
	\right),
}
where $\phi$ is the azimuthal angle, and $\theta$ the polar angle.

\subsubsection*{Exchange interaction}
The exchange interaction actually gives rise to ferromagnetism and originates from the Pauli repulsion between two neighboring spins. For ferromagnets it is energetically favorable for the two spins to align because of this interaction and it is therefor also called the exchange stiffness. A misalignment of neighbors gives an energy contribution 
\eq{
	-J_\textrm{s} \mathbf{\Omega}(\mathbf{x})\cdot\mathbf{\Omega}(\mathbf{x}+\mathbf{a})=-J_\textrm{s}\cos(\d\omega),
}
where $J_\textrm{s}$ is the spin exchange stiffness, $\mathbf{a}$ the vector connecting two neighbors and $\d\omega$ is the angle between the two magnetizations. In the continuum limit we express this angle in terms of the gradients of $\q$ and $\f$ by $\d\omega\approx|a|\sqrt{(\nabla\q)^2+\sin^2(\q)(\nabla \f)^2}$. Now we sum over all nearest neighbor sites, which in the continuum limit becomes an integral. By expanding the cosine to second order and defining the continuum exchange interaction $J=J_\textrm{s} a^{-1}$ with $a$ now the lattice spacing we obtain 
\eq{
	\label{eq:exchange1}
	E_\textrm{exchange}(\q,\f)=\frac{J}{2}\int d^3\mathbf{x}\left((\nabla\q)^2+\sin^2(\q)(\nabla\f)^2\right).
}
A typical value for $J$ is around $10^{-11}$ Jm$^{-1}$ and for $a$ is around $3$ \r{A}.

\subsubsection*{Perpendicular magnetic anisotropy}
Phenomenologically, we introduce an anisotropy energy for an easy $z$-axis, by
\begin{equation}
\label{eq:anisotropy}
E_\textrm{PMA}(\q,\f)=-\frac{K_\textrm{P}}{2}\int d^3\mathbf{x}\cos^2(\q).
\end{equation}
The subscript $P$ stands for PMA. From symmetry arguments higher order terms in $\f$ and $\q$ could also be included, but we omit them for simplicity. It will be convenient to define an effective anisotropy magnetic field strength $H_\textrm{P}=K_\textrm{P}/M_\textrm{S}$ with $M_\textrm{S}$ the saturation magnetization. It should be noted that $H_\textrm{P}$ is only introduced for notational convenience and cannot be physically interpreted as an effective magnetic field. $H_\textrm{P}$ has typical values around $0.1$ T and $M_\textrm{S}$ around $5\times 10^5$ Am$^{-1}$.

\subsubsection*{Demagnetization}
Dipole-dipole interactions also contribute to the energy and prefer Bloch DWs. The dipole-dipole potential energy is written as a Zeeman energy from the demagnetization field $\mathbf{H}_\textrm{d}$
\eq{
	E_\textrm{d}=-\frac{1}{2}\int d^3\mathbf{x}\mathbf{M}(\mathbf{x})\cdot\mathbf{H}_\textrm{d}(\mathbf{x}).
}
Following Tarasenko \textit{et al.}\cite{Tar98}, we perform a Fourier transform of the demagnetisation field, its corresponding magnetic potential $\Phi$ (such that $\mathbf{H}_\textrm{d}=\nabla\Phi$) and the magnetization $\mathbf{M}=M_\textrm{S}\mathbf{\Omega}$ to obtain an expression for the above energy in terms of the fourier modes using Maxwel's equation 
\eq{
	\nabla\cdot \mathbf{B}=\nabla \cdot (\mathbf{H}+4\p\m_0\mathbf{M})=0;
}
we thus find
\eq{
	E_\textrm{d}=\frac{\m_0}{4\p^2}\int d^3\mathbf{k}\frac{1}{k^2}(\mathbf{M_k}\cdot\mathbf{k})(\mathbf{M_{-k}}\cdot\mathbf{k}).
}
Here, $k=|\mathbf{k}|$. Now we perform the inverse Fourier transform of the $z$ component. 
\eq{
\mathbf{M_k}=\int dz \mathbf{M}_\k(z)e^{-i k_z z},
}
where $\bm{\k}=(k_x,k_y)^T$. This results in 
\ea{
	E_\textrm{d}=&\frac{\m_0}{4\p^2}\int d^2\bm{\k}\int dz\int dz'\int dk_z\frac{1}{\k^2+k_z^2}e^{-ik_z(z-z')}\\ \nonumber
	(&\mathbf{M}_{\bm{\k}}(z)\cdot\bm{\k}+M_{\bm{\k}}^z(z)k_z)(\mathbf{M_{-\bm{\k}}}(z')\cdot\bm{\k}+M_{-\bm{\k}}^z(z')k_z).
}
We take out an effective contribution to the easy $z$-axis anisotropy by using:
\ea{
\int d^3\mathbf{x}M_z^2=&\int dz\int \frac{d^2\bm{\k}d^2\mathbf{\k'}}{(2\p)^4}M_{\bm{\k}}^zM_\mathbf{\k'}^ze^{i\mathbf{x}\cdot(\bm{\k}+\mathbf{\k'})}\\
=&\int dz\int \frac{d^2\bm{\k}}{(2\p)^2}M_{\bm{\k}}^zM_\mathbf{-\k}^z. \nonumber
}
Then we evaluate the integral over $k_z$ and are left with 
\ea{\label{eq:dme2}
	E_\textrm{d}=&\frac{\m_0}{4\p}\int d^2\bm{\k}\int dz\int dz'e^{-\k|z-z'|}\\ \nonumber
	&\left(\frac{1}{\k}(\mathbf{M}_{\bm{\k}}(z)\cdot\bm{\k})(\mathbf{M_{-\bm{\k}}}(z')\cdot\bm{\k})-\k M_{\bm{\k}}^z(z)M_{-\bm{\k}}^z(z')\right.\\ \nonumber
	&\left.-2i \text{ sign}(z-z')(\mathbf{M}_{\bm{\k}}(z)\cdot\bm{\k})M_{-\bm{\k}}^z(z')\right).
}
The first term is the energy of the magnetic field declination away from the DW. The second term is the energy associated with the change in magnetization at the boundaries. The third term describes the twisting of the DW. \cite{Mal79,Hub74} 

Next, under the assumption that the film thickness $t$ is small (i.e. $t<\lambda$), the magnetization is homogeneous along the $z$ direction and we evaluate the integrals over $z$ and $z'$. We use
\ea{
\int_{-t/2}^{t/2}dz\int_{-t/2}^{t/2}dz'e^{-|z-z'|\k}&=\frac{2t}{\k}\left(1-\frac{1-e^{-\k t}}{\k t}\right)\\
&\approx t^2 \text{, for } \k t\ll1,\nonumber
}
and
\eq{
	\int_{-t/2}^{t/2}dz\int_{-t/2}^{t/2}dz'e^{-|z-z'|\k}\text{ sign}(z-z')=0,
}
by symmetry. Furthermore, we rotate the IP components of $\mathbf{M}$ such that the $x$ component aligns with the DW normal. This means $\mathbf{\Omega}(\f,\q)\rightarrow\mathbf{\Omega}(\f-\a,\q)$, where $\a$ is the angle (w.r.t. the x-axis) of the DW normal. And by translation symmetry the field is also homogeneous in the $y$ direction. That means
\ea{
	\mathbf{M}_{\bm{\k}}\cdot\bm{\k}=M^x_{\bm{\k}}\text{, and}\\
	M^x_{\bm{\k}}=2\p\d(k_y)M^x_{k_x}.
}
For the DW dynamics only the first term in \cref{eq:dme2} is relevant. As we shall see from the DW profile in the following, $M_{k_x}$ converges rapidly to 0 for increasing $k_x$, hence for small $t$, we assume $\k t\ll1$. Then the demagnetization energy is reduced to 
\eq{
	\label{eq:dme3}
	E_\textrm{d}\approx\frac{t^2\m_0}{2}\int dk |k|M^x_kM^x_{-k}\int dy.
}
The integral over $y$ indicates that the demagnetization energy increases linearly with the DW length. Once we know the DW profile (i.e. $\q$ and $\f$ as a function of $x$), we then explicitly compute the inverse Fourier transform
\eq{
	\label{eq:mkx}
	M^x_k&=\int_{-\infty}^{\infty} dx M_x e^{ikx}\\ \nonumber
	&=M_\textrm{S}\int_{-\infty}^{\infty} dx \sin(\q)\cos(\f-\a) e^{ikx},
}
to obtain the final expression for the demagnetization energy by working out all the integrals.
\subsubsection*{Zeeman energy}
The energy contribution of the external magnetic field $\mathbf{H}$ is the Zeeman energy, given by
\ea{
	\label{eq:ext}
	E_\textrm{external}(\q,\f)&=M_\textrm{S}\int d^3\mathbf{x}(-\mathbf{\Omega}(\mathbf{x})\cdot\mathbf{H})\\ \nonumber
	&=-M_\textrm{S}\int d^3\mathbf{x}\left(H_x \sin(\q)\cos(\f)\right.\\ \nonumber
	&\left.\qquad+H_y\sin(\q)\sin(\f)+H_z\cos(\q)\right).
}

\subsubsection*{Dzyaloshinskii-Moriya interaction}
For convenience we rotate our axes such that the gradient of the magnetization $\omega$ points along the $x$ axis and the interface between the ferromagnet and (heavy spin orbit coupling) normal metal is perpendicular to the $z$ axis. For two neighboring spins along the $x$ direction, the energy contribution due to the interfacial Dzyaloshinskii-Moriya interaction (DMI) is given by
\begin{equation}
E_\textrm{DMI}=\mathbf{D}_\textrm{s}\cdot(\mathbf{\Omega}(\mathbf{x})\times\mathbf{\Omega}(\mathbf{x}+a\hat{x})),
\end{equation}
where $\mathbf{D}_\textrm{s}=D_\textrm{s}\hat{y}$ is the DMI. From Ref.~[\onlinecite{Fer80}] we compute $D_\textrm{s}$ to be in te order of $10^{-23}$ J. We approximate the magnetization direction of the neighboring particle ($\mathbf{\Omega}(\mathbf{x}+a\hat{x})$) in terms of $\mathbf{\Omega}(\mathbf{x})$ by $\mathbf{\Omega}(\mathbf{x}+a\hat{x})\approx\mathbf{\Omega}(\mathbf{x})+a\frac{\partial \mathbf{\Omega}(\mathbf{x})}{\partial x}$. Then, summing over all sites and taking the continuum limit, we get
\eq{
	E_\textrm{DMI}=D_\textrm{s}\int \frac{dy}{a} \int dx (\hat{y}\cdot(\mathbf{\Omega}(x)\times\frac{\partial \mathbf{\Omega}}{\partial x}).
}
Where we used that the system is homogeneous along the $y$ direction.
We then define $D=D_\textrm{s} a^{-1} t^{-1}$ to obtain an interfacial DMI energy per DW area. $D$ has typical values around $10^{-4} Jm^{-2}$. We also define an effective DMI field $H_\textrm{D}=D/(M_\textrm{S} \l)$. In the final section of this supplemental material we apply our model to DW velocity measurements of two series of samples with varying thickness to extract the value of $H_\textrm{D}$ of these samples and verify the proportionality $H_\textrm{D}\propto t^{-1}$.

Inserting the definition of $\Omega$ and working out the cross and inner product, we obtain in terms of the spherical coordinates $\theta$ and $\phi$.
\begin{equation}
E_\textrm{DMI}(\q,\f)=t M_\textrm{S} \l H_\textrm{D}\int dx dy \cos(\f)\frac{\6\q}{\6 x}.
\end{equation}
In this case $H_\textrm{D}$ is not only introduced for notational convenience, but also has the physical interpretation of a local IP magnetic field directed along $-\nabla\q$. 
Let $\a$ be the angle between the DW normal and the applied external IP magnetic field. Then we rotate our axes back such that the $x$ axis lied parallel to the IP magnetic field. This effectively shifts $\f\rightarrow\f-\a$.

\subsection{DW profile}
Now we derive an expression for $\q$ in equilibrium for a system with no external magnetic field, nor DMI, nor Bloch anisotropy. Then $\f$ is constant in equilibrium. We then apply the Euler-Lagrange equation to the energy density, which is distilled from the above.
\ea{
	\label{eq:ELt1}
	\frac{\partial\mathcal{E}}{\partial\theta}=\nabla\frac{\partial\mathcal{E}}{\partial(\nabla\theta)} \text{, so}
	\\
	\label{eq:ELt2}
	-\frac{K_\textrm{P}}{2}\frac{\partial\cos^2(\theta)}{\partial\theta}=J\nabla^2\theta.
}
For this system, derivatives with respect to the $z$ coordinate are negligible and we align the $y$-axis with the DW, so $\nabla\theta=\frac{\partial\theta}{\partial x}$. Now if we multiply \cref{eq:ELt2} with $\frac{\partial\theta}{\partial x}$ and integrate over $x$, we find:
\begin{equation}
-\frac{K_\textrm{P}}{J}\cos^2(\theta)+C=\left(\frac{\partial\theta}{\partial x}\right)^2.
\end{equation}
With $C$ and integration constant. As we send $x\rightarrow-\infty$ we know that $\frac{d\theta}{dx}\rightarrow0$ and $\cos(\theta(x))\rightarrow Q=\pm 1$. $Q$ is the charge of the DW distinguishing between an up-down ($+1$) and down-up ($-1$) wall. So we find $C=\frac{K_\textrm{P}}{J}$. Using the goniometric identity $\sin^2a+\cos^2a=1$ and taking the square root we find:
\begin{equation}
Q \frac{\partial \theta}{\partial x}=\sqrt{\frac{K_\textrm{P}}{J}}\sin(\theta).
\end{equation}
Now we separate variables. A primitive of the function $\frac{1}{\sin(\theta)}$ is $\ln(\tan(\frac{\theta}{2}))$. To find the integration constant we define the DW position $x_\textrm{DW}$ as such that $\theta(x=x_\textrm{DW})=\pi/2$. And so we find:
\begin{equation}
\ln\left(\tan(\theta/2)\right)=\frac{Q}{\lambda}(x-x_\textrm{DW}),
\end{equation}
with $\lambda=\sqrt{\frac{J}{K_\textrm{P}}}$ a typical length scale for the DW width. We solve for $\q$ to find
\begin{equation}
\label{eq:theta1}
\theta(x)=2\arctan\left(e^{\frac{Q}{\lambda}(x-x_\textrm{DW})}\right).
\end{equation}

\subsection{DW profile for tilted domains}
An IP magnetic field also changes the orientation of the magnetic moment inside the domains. For simplicity we assume $Q=+1$. The equilibrium orientation has to balance PMA and the IP magnetic field, yielding a different internal magnetization profile. The energy density well inside the domain is given by
\eq{
	\label{eq:E4}
	\mathcal{E}_\textrm{in}=-M_\textrm{S}\left(\mathbf{H}_\textrm{IP}\cdot\hat{m}\sin(\theta)+\frac{H_\textrm{PMA}}{2}\cos^2(\theta)\right).
}
Here $H_\textrm{P}=K_\textrm{P}/M_\textrm{S}$ is the effective magnetic field of the PMA, $\mathbf{H}_\textrm{IP}$ is the IP vector of the applied magnetic field and $\hat{m}(\f)$ the IP vector of the internal magnetization. We minimize this energy, where we assume $H_\textrm{P}\gg| \mathbf{H}_\textrm{IP}\cdot\hat{m}|$, because else there will be no perpendicularly magnetized domains anymore. The two solutions are 
\ea{
	\label{eq:theta}
	\theta=\arcsin(\b), \qquad \text{or}\qquad \theta=\pi-\arcsin(\b),\\
	\text{with} \qquad \b=\frac{\mathbf{H}_\textrm{IP}\cdot\hat{m}}{H_\textrm{PMA}}.
}
Which precisely characterizes the magnetization in the the up and down domains respectively. We have the bounds $|\b|\ll1$. 

This immediately yields an effect on the driving force as the Zeeman energy difference now is proportional to
\ea{
	\label{eq:DZE}
	\D E_\textrm{Zeeman}&\propto \frac{1}{2}H_z(\cos(\pi-\theta)-\cos(\theta))\\ \nonumber
	&=H_z\cos(\theta)\\ \nonumber
	&=H_z\sqrt{1-\b^2}.
}

We investigate how these new boundary conditions affect the energy density. Accounting for IP magnetic fields and DMI, the energy density reads
\ea{
	\label{eq:edt}
	\mathcal{E}=\frac{J}{2}(\nabla\q)^2&+M_\textrm{S}\Big(-\frac{H_\textrm{P}}{2}\cos(\q)^2-\mathbf{H}_\textrm{IP}\cdot\hat{m}\sin(\q)\\ \nonumber
	&+H_\textrm{D}\l\cos(\f-\a)\nabla\q\Big).
}
The Euler-Lagrange formalism yields
\eq{
	J\6_x^2\q=M_\textrm{S}\left(H_\textrm{P}\cos(\q)\sin(\q)-\mathbf{H}_\textrm{IP}\cdot\hat{m}\cos(\q)\right).
}
We divide out $M_\textrm{S} H_\textrm{P}$ and use the definitions of $\l$ and $\b$ to rewrite the above as
\eq{
	\l^2\6_x^2\q=\cos(\q)\left(\sin(\q)-\b\right).
}
As we can see, the DMI has no effect on the DW profile because it depends only linearly on $\nabla\q$. Using $2(\6_x\q) (\6_x^2\q)=\6_x\left((\6_x\q)^2\right)$ and $2(\6_x\q)\cos(\q)\sin(\q)=\6_x \sin(\q)^2$, the above yields
\eq{
	\left(\l\6_x\q\right)^2=\left(\sin(\q)^2- 2\b\sin(\q)\right)+C,
}
where $C$ is a constant which we now determine from the boundary conditions. As $x\rightarrow\infty$, $\6_x\q\rightarrow0$ and $\sin(\q)\rightarrow\b$,  we obtain:
\eq{
	C=\b^2.
}
Thus
\eq{
	\label{eq:eomt}
	(\l\6_x\q)^2=\left(\sin(\q)-\b\right)^2\text{ , that is}\qquad\l\frac{\6\q}{\6x}=\sin(\q)-\b.
}
With boundary condition $\q(x_\textrm{dw})=\p/2$, we solve this equation and obtain 

\begin{widetext}
	\ea{
		\label{eq:thx}
		\theta(x)&=2\arctan\left(\frac{1}{\b}-\b\sqrt{1-\b^2}\tanh\left(\frac{(x-x_\textrm{dw})}{\l}\frac{\sqrt{1-\b^2}}{2}-\text{arccoth}\left(\frac{1+\b}{\sqrt{1-\b^2}}\right)\right)\right);\\
		&=2\arctan\left(e^{\frac{x-x_\textrm{dw}}{\lambda}}\right)-\b\tanh\left(\frac{x-x_\textrm{dw}}{\l}\right)+\mathcal{O}(\b^2).
	}
\end{widetext}

\begin{figure}[h]
	\center
	\includegraphics[width=\linewidth]{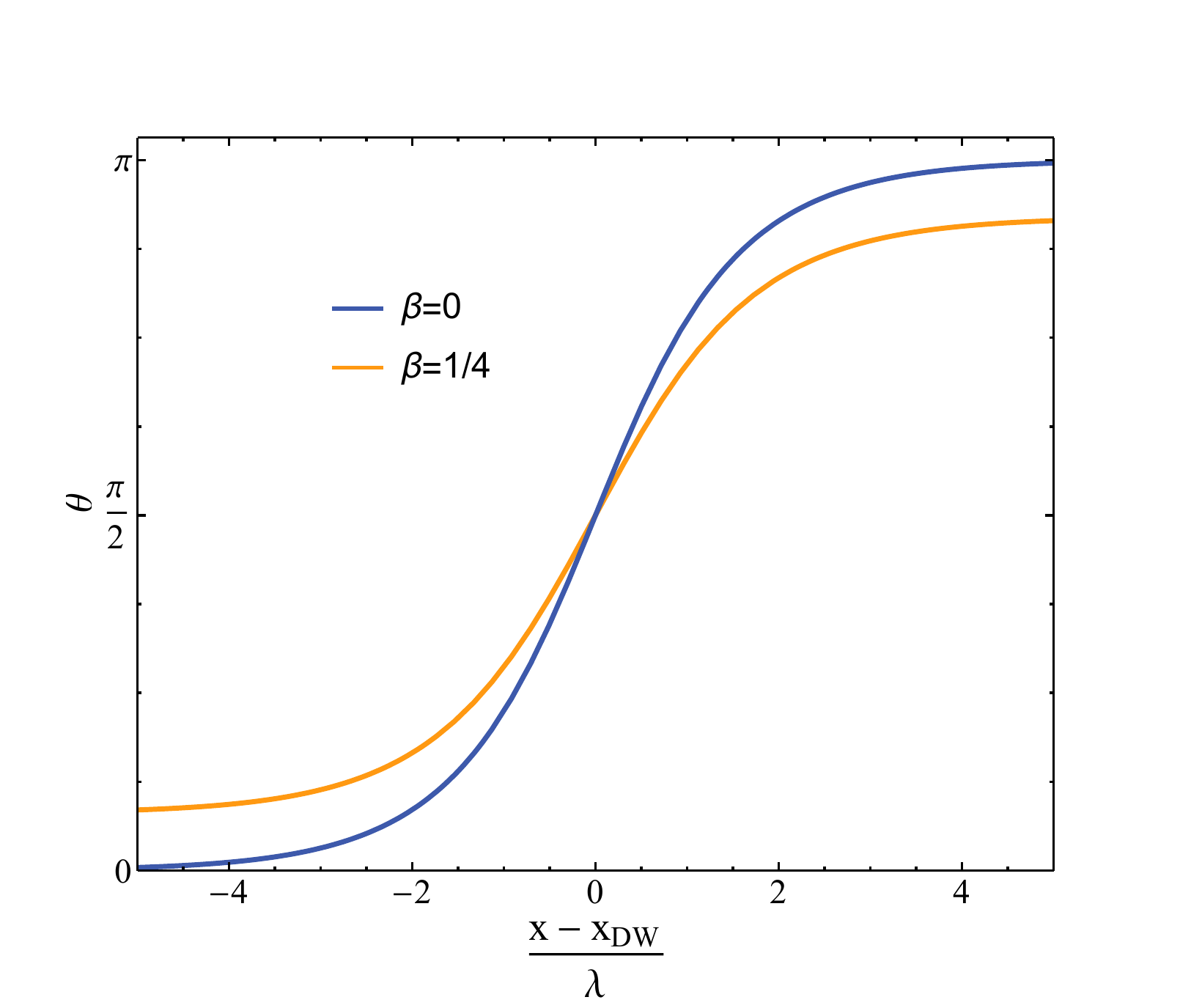}
	\caption{\label{fig:dwshape} Effect of IP magnetic field on the DW profile parameter $\th$. A side view of the corresponding magnetization is illustrated in Fig. 1 (c) of the main article.}
\end{figure}

This DW shape is shown in \cref{fig:dwshape} for $\b=1/4$ along with the $\b=0$ case. The result leaves the DW thickness $\l$ intact, as can also be seen in the figure. 


\subsection{Energy density}
We now assume that along the $x$-axis $\f$ is constant. Then we integrate our energy density over $x$ using \cref{eq:thx} (or \cref{eq:theta1} for the untilted case), subtracting the energy density of a system with a DW at $x_\textrm{dw}=0$, and we are left with an equilibrium (i.e. $H_z=0$) energy density. We work out the case for the demangetization energy explicitly based on the work of Tarasenko \textit{et al.}\cite{Tar98}. We start out by reproducing the known results for $\b=0$. 

For convenience we define $X=(x-x_\textrm{dw})/\l$, so $dx=\l dX$. Note that $\sin(\q)=1/\cosh(X)$, which has poles at $X_n=\frac{\p i}{2}(2n+1)$ for $n\in\mathds{Z}$. We now evaluate \cref{eq:mkx} by closing the contour in the upperhalf of the complex plane along an infinity modulus semi circle. Due to the exponent, this contribution vanishes, and we use the residue theorem. The residue of the integrand at $X_n$ is given by 
\eq{ 
	\frac{e^{\frac{1}{2}\p k \l(2n+1)}}{i\cos(n\p)}.
}
Now we sum over all poles in within the contour (i.e. $n>0$) to find
\eq{
	M_k^x=M_{-k}^x=M_S \cos(\f-\a) \frac{\p\l}{\cosh(\frac{k\p\l}{2})}.
}
Which indeed vanishes rapidly for increasing $k$. Now we evaluate the integral over $k$ in \cref{eq:dme3} to obtain the energy density (per sample thickness per DW length)
\eq{
	\mathcal{E}_d=4 \ln(2) t \m_0 M_\textrm{S}^2 \cos^2(\f-\a).
}
Hence we define an effective Bloch anisotropy field (referring to the fact that this energy contribution favors a Bloch type DW) $H_\textrm{B}=4 \ln(2) \frac{t \m_0 M_\textrm{S}}{\p\l}$.

In total, the energy density is given by
\begin{widetext}
\ea{
	\label{eq:enden2}
	\mathcal{E}(\a,\f)&=2\sqrt{J K_\textrm{P}}+4\m_0M_\textrm{S}^2 t\ln(2)\cos^2(\f-\a)-M_\textrm{S}\p\l(H_x\cos(\f)+H_y\sin(\f))+\p D\cos(\f-\a)\\ \nonumber
	&=2\frac{J}{\l}+M_\textrm{S} \p \l \left(H_\textrm{B}\cos^2(\f-\a)-(H_x\cos(\f)+H_y\sin(\f))+H_\textrm{D}\cos(\f-\a)\right)
}
\end{widetext}

When $\f$ varies along the DW, the energy density obtains a bending contribution following from the exchange interaction:
\eq{
	\label{eq:eben}
	\mathcal{E}_\textrm{bend}= 2 J\l \left(\frac{\6\f}{\6 s}\right)^2.
}

\begin{figure}[ht!]
	\includegraphics[width=\linewidth]{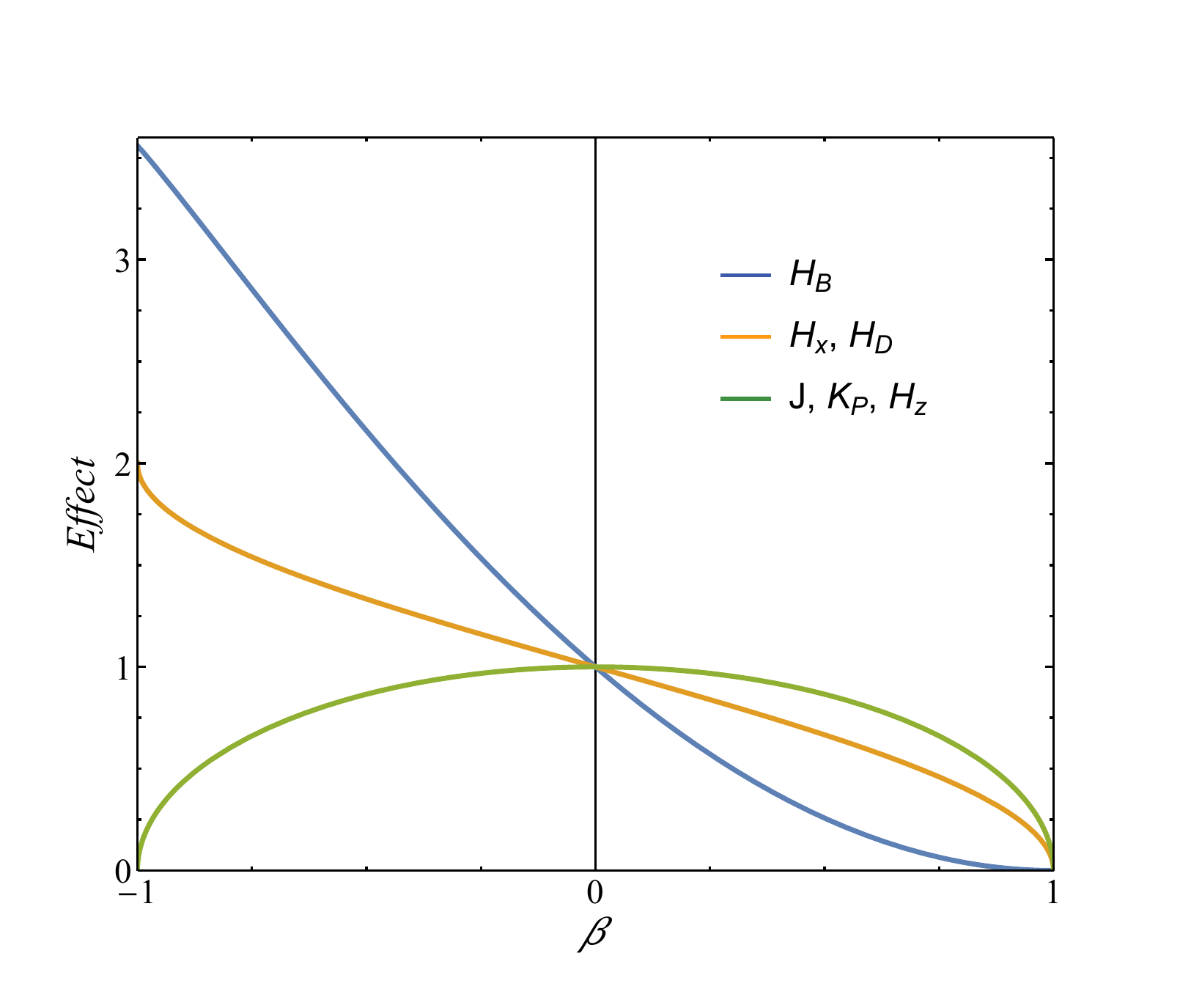}
	\caption{\label{fig:Em} The effect of the $\b$ parameter on $H_\textrm{B}$ (blue), $H_\textrm{D}$ and $H_x$ (orange), $J$, $K_\textrm{P}$ and $H_z$ (green), normalized to their respective values for $\b=0$.}
\end{figure}

Now we calculate the same energy density for $\b\neq0$. The more elegant \cref{eq:eomt} is used to simplify these integrals. For convenience we first define
\ea{
	\label{eq:fb}
	f(\b)=&\frac{4}{\p}\left(\arcsin(\sqrt{\frac{1+\b}{2}})-\arcsin(\b)\right)\\ \nonumber
	\approx&1-\frac{2}{\p}\b+\mathcal{O}(\b^3). \\
	\label{eq:gb}
	g(\b)=&\frac{(1-\b^2)}{4\ln(2)}\left(-H_{-b}-H_{b}+b(\psi^{(1)}(1-b)-\psi^{(1)}(1+b))\right)
} 
The results are:
\begin{itemize}
		\item Exchange interaction
		\eq{
			E_\textrm{Exchange}(\f)=
			\frac{J}{\l}\left(\sqrt{1-\b^2}-\frac{\p\b}{2}f(\b)\right).
		}
		\item PMA
		\eq{
			E_\textrm{PMA}(\f)=
			K_\textrm{P}\l\left(\sqrt{1-\b^2}+\frac{\p\b}{2}f(\b)\right).
		}
		Using the definition of $\l$ we combine the two above expressions, cancelling the inverse cosine functions. The result is $2\frac{J}{\l}\sqrt{1-\b^2}$.
		\item Bloch anisotropy\\
		Using the residue theorem we compute $M_k^x$ so that the Bloch anisotropy energy is now given by
		\eq{
			E_\textrm{Bloch}(\f)&=4\m_0M_\textrm{S}^2 t\p^2\l^2\cos^2(\f-\a)\times \\
			&\qquad \int_{-\infty}^\infty dk \frac{\sinh^2\left(\frac{k \l \arccos(\b)}{\sqrt{1-\b^2}}\right)}{\sinh^2\left(\frac{k \l \p}{\sqrt{1-\b^2}}\right)}|k|\\ \vspace{3mm}
			&=8\m_0M_\textrm{S}^2 t\cos^2(\f-\a) (1-\b^2) \times \\
			&\qquad\int_{0}^\infty dK \frac{\sinh^2\left(K b\right)}{\sinh^2\left(K\right)}|K|.
		}
		We have changed variables $K=\frac{k\p\l}{\sqrt{1-\b^2}}$ and defined $b=\arccos(\b)/\p$, $0\leq b\leq 1$. Now we work out the integral using the geometric series \cite{MF}
		\eq{
			\sum_{n=0}^{\infty}(n+1)e^{-2nK}=\frac{1}{(1-e^{-2K})^2}.
		}
		We write out the definition of hyperbolic sine in terms of exponentials and use
		\eq{
			\int_{0}^{\infty}dK K e^{-a K}=\frac{1}{a^2},
		}
		for $a>0$, to evaluate the integral. Finally, we evaluate the sum over $n$ to find
		\eq{
			E_\textrm{Bloch}(\f)=8\m_0M_\textrm{S}^2 t\cos^2(\f-\a) \ln(2)\frac{(1-\b^2)}{4\ln(2)} \times \\
			\left(-H_{-b}-H_{b}+b(\psi^{(1)}(1-b)-\psi^{(1)}(1+b))\right).
		}
		$H_{z}$ is the Harmonic number of $z$ and $\psi^{(1)}$ is the first derivative of the digamma function. When $\b=0$, $b=1/2$ and $2+H_{-1/2}=H_{1/2}=2-2\ln(2)$ and $\psi^{(1)}(3/2)=\psi^{(1)}(1/2)-4=\p^2/2$, so we indeed obtain the known limit. This result is shown in \cref{fig:Em} as a function of $\b$.
		
		\item Zeeman energy
		\eq{
			E_\textrm{Zeeman}(\f)=-M_\textrm{S}\l\p(H_x\cos(\f)+H_y\sin(\f))f(\b)
		}
		\item DMI
		\eq{
			E_\textrm{DMI}(\f)=4\p D\cos(\f)f(\b)
		}
	\end{itemize}
	
We absorb this effect by a redefinition of the effective fields and parameters as follows
\begin{itemize}
	\item $J\rightarrow J\sqrt{1-\b^2}$;
	\item $K_\textrm{P}\rightarrow K_\textrm{P}\sqrt{1-\b^2}$, which keeps $\l$ fixed;
	\item $H_\textrm{IP}\rightarrow H_\textrm{IP}f(\b)$;
	\item $H_{z}\rightarrow H_{z}\sqrt{1-\b^2}$;
	\item $H_\textrm{D}\rightarrow H_\textrm{D}f(\b)$;
	\item $H_\textrm{B}\rightarrow H_\textrm{B} g(\b)$.
\end{itemize}

In \cref{fig:Em} the different effects of $\b$ are plotted, normalized with respect to the $\b=0$ value.

\begin{figure}
	\includegraphics[width=\linewidth]{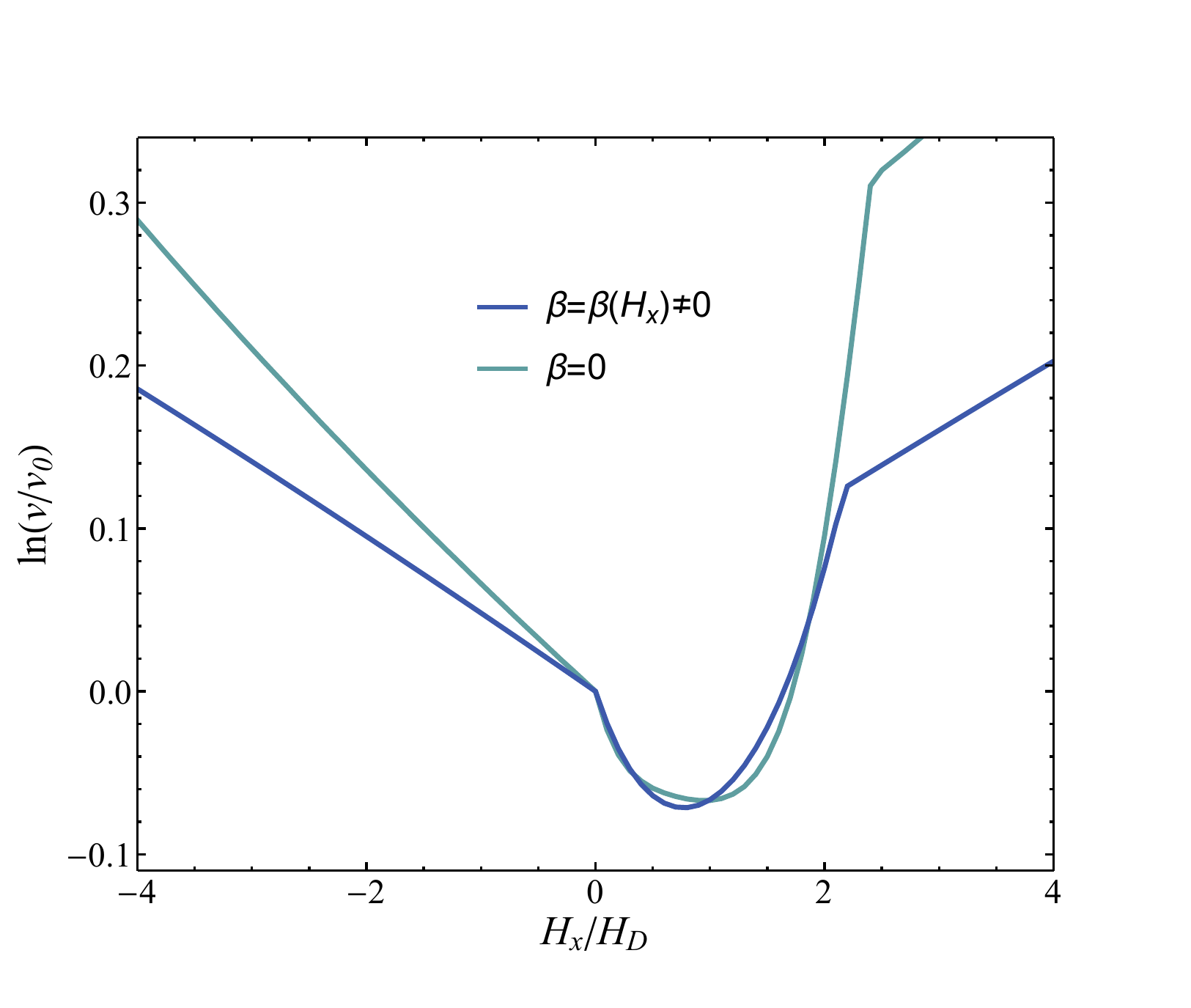}
	\caption{\label{fig:vnb}The DW velocity corrected for tilting of the magnetic domains (blue) compared with the model where this effect is neglected and $\b$ is set to $0$ (light blue). The effect is significant and therefore implemented in our model.}
\end{figure}

The effect of this domain tilting on the final DW velocity as a function of $H_x$ is shown in \cref{fig:vnb} for typical parameter values (see \cref{tab:cnp}).

\clearpage
\section{Creep theory in depth}
\label{app:B}
In this section we work out the theory of creep for magnetic DWs, based on the review by Blatter \textit{et al.} who used creep theory to describe the motion of magnetic vortices in superconductors \cite{Bla94rev}. To start, we follow the analytic derivation by Lemerle \cite{Lem98}, where no IP magnetic field is applied and expose relevant details to be used later on. By assuming the elasticity to be non dispersive, the results of Je \textit{et al.} immediately follow \cite{Je13}. However, when the elasticity is considered to be dispersive, one has to abandon the analytic approach of Lemerle and use numerical optimizations to obtain the DW velocity. We conclude this section by commenting on the work by Pellegren et al \cite{Sok17}, who, instead of optimizing actually keep some parameters to fit; we reproduce their work but do optimize.

Creep theory attempts to describe  some manifold in a pinning potential, whose barrier height is larger than the driving force, but still exhibits a net macroscopic motion of due to fluctuations (e.g. quantum or thermal). The rate at which these fluctuation occur is computed from the Arrhenius law \cite{Arr89a,Arr89b}; a deformation due to a fluctuation costs elasticity energy, but yields energy from the driving force. These costs and yield scale differently with the size of the fluctuation so that there is a optimal size $L_\textrm{opt}$ above which the deformation can grow. At this optimal deformation length the energy cost is maximal and equal to the energy barrier the nucleation problem has to overcome. 

Hence the recipe is clear: Determine the elasticity energy cost and the driving energy gain as a function of the deformation length $L$. Subtract the two and optimize for $L$. As we shall see for a non dispersive stiffness, this optimization can be done analytically and yields the famous creep law $\ln(v)\propto-H_z^{-1/4}$. In fact, this law has become so famous that the exact proportionality has become a measure of how well a systems dynamics are in the creep regime. We show that a purely theoretical creep model also exhibits deviations from this creep exponent.

\subsection{Non dispersive energy density}
We consider a straight DW and deform a segment of length $L$ with amplitude $u$ as is shown in the figure of the main article (Lermerle considers a segment of length $2L$) \cite{Lem98}. The magnetization to the left of the DW is parallel to the external driving field. Hence the deformation is energetically favorable with respect to this external field due to the Zeeman energy. However, due to the deformation, the DW gets stretched, which is energetically unfavorable. Furthermore, we assume that the disorder is expressed trough a pinning potential that has some characteristic length scale. We assume $u$ to be small, e.g. the increase in the DW length due to the deformation is expressed as $2\sqrt{u^2+(L/2)^2}-L\approx 2u^2/L$. There is no external IP magnetic field, nor do we consider DMI, nor dipole-dipole interaction. In summary, the energy difference between the deformed DW and the straight DW is given by:
\eq{
	\label{eq:E1}
	\D E(u,L)=\mathcal{E}_\textrm{el}\frac{2u^2}{L}-\sqrt{\D\xi^2L}-M_\textrm{S} H_z t \frac{L u}{2}. 
}
Here, $\mathcal{E}_\textrm{el}$ is the elastic energy density, $t$ is the film thickness, $\l$ the DW width, $J$ is the exchange stiffness, $K$ is the effective anisotropy energy, $\D$ is the pinning strength scaling, $\xi$ is the characteristic length scale of the pinning potential, $M_\textrm{S}$ is the saturation magnetization (in Tesla) and $H_z$ is the applied out of plane (OOP) magnetic field (in Oersted Oe=A/m). The three terms in the RHS of the above equation thus account for stretching, pinning and Zeeman energy. The square root in the pinning term reflects that only fluctuations in the pinning potential affect the DW dynamics.

The elasticity is determined by subtracting the energy of an unperturbed DW from the energy of the deformed DW. When the energy density is non dispersive, this difference is just the energy density times the DW elongation. However, the elongation is a function of both $u$ and $L$. So we now set out to express $u$ in terms of $L$ using the exponential wandering relation.

We define a length scale $L_\textrm{c}$ for the DW segment length, above which the DW can adjust itself elastically, so that locally it is in the optimal configuration with respect to the pinning potential. This length scale is determined by the optimal balance between elasticity and pinning energy, for a deformation amplitude of $u=\xi$ (i.e. a metastable state which has hopped one pinning site). This yields
\eq{
	L_\textrm{c}=2(2\mathcal{E}_\textrm{el}^2\xi^2/\D)^{1/3}.
}
We also define a critical field strength $H_\textrm{c}$ as the external OOP field strength required to move the DW with $L=L_\textrm{c}$ and $u=\xi$ in the absence of thermal fluctuations. It is given by
\eq{
	H_\textrm{c}=\frac{4\mathcal{E}_\textrm{el}\xi}{L_\textrm{c}^2M_\textrm{S}t},
}
which is found by equating the pinning energy and the Zeeman energy with $L=L_\textrm{c}$ and $u=\xi$.

The spatially and thermally averaged correlation function $\la\la (u(y+L)- u(y))^2\ra\ra$ is proportional to $(L/L_\textrm{c})^{2\z}$, where $\z$ is the characterizing wandering exponent. For 1D DWs it has been proven that $\z=2/3$ \cite{Kar85,Fis91rev}. We thus assume a scaling law for the displacement 
\eq{
	\label{eq:wex}
	u=u_\textrm{c}(L/L_\textrm{c})^\z,
}
where $u_\textrm{c}$ is a transverse scaling parameter. We insert this relation into \cref{eq:E1} and drop out the pinning energy to find
\eq{
	\label{eq:E2}
	\D E(L)=\frac{2\mathcal{E}_\textrm{el} u_\textrm{c}^2}{L_\textrm{c}}\left(\frac{L}{L_\textrm{c}}\right)^{2\z-1}-\frac{1}{2}M_\textrm{S} H_z t u_\textrm{c} \left(\frac{L}{L_\textrm{c}}\right)^{\z+1}. 
}
When we maximize (notice that Lemerle calls this a minimization) this expression for $L$ we find 
\eq{
	\label{eq:Lmin}
	L_\textrm{opt}=L_\textrm{c}\left(\frac{\z+1}{2\z-1}\frac{M_\textrm{S} H_ztL_\textrm{c}^2}{4u_\textrm{c}\mathcal{E}_\textrm{el}}\right)^{\frac{1}{\z-2}}.
}
Now we insert $\z=2/3$ and \cref{eq:Lmin} into \cref{eq:E2} to find the energy barrier
\eq{
	\label{eq:E3}
	F_\textrm{b}=\frac{4}{5^{5/4}}\frac{u_\textrm{c}^{9/4}\D^{1/2}}{\xi}\left(\frac{\mathcal{E}_\textrm{el}}{H_z M_\textrm{S} t}\right)^{1/4}.
}
By Arrhenius' equation the hopping rate is proportional to the exponential of the Boltzman weighted energy barrier. The DW velocity is found because it should be proportional to this hopping rate. In conclusion, we obtain the following proportionality
\eq{
	\label{eq:v1}
	ln(v)\propto -\left(\frac{\mathcal{E}_\textrm{el}}{H_z}\right)^{1/4}.
}
Notice that we only used that $\mathcal{E}_\textrm{el}$ is the proportionality constant of the energy gain from the deformation, proportional to $2u^2/L$. As soon as $\mathcal{E}_\textrm{el}$ can no longer be approximated with the above proportionality. So next, we consider a dispersive energy density and adjust the above derivation accordingly.

\subsection{Dispersive energy density}
There are three components that make up elasticity. Most familiar is stretching: There is an DW energy cost per length, and a deformation elongates the DW. To minimize this energy cost, the DW is kept short and thus straight. Some systems (including ours) need to distinguish between stretching and bending; although both try to keep the manifold straight the underlying mechanisms are different. In our system both stretching and bending energy costs are caused by the exchange interaction. A third elasticity component is formed by orientation. In our system this component starts playing a role when applying an external field. For example a Bloch DW becomes more flexible when applying an IP magnetic field perpendicular to the DW normal as the magnetic moment at the DW will then be more aligned with the IP magnetic field upon any deformation. Similarly a Néel wall will become stiffer in the same situation.

So suppose in general we have some DW energy density $\mathcal{E}(\f,\a)$ which is a function of the azimuthal angle of the magnetization of the sample at the DW $\f$, and the angle $\a$ of the DW normal w.r.t. the applied in-plane magnetic field. After minimizing it for $\f$ this energy $\mathcal{E}_\textrm{min}(\a)$ is only a function of $\a$. We assume a DW deformation that has some profile $\a(s)$ parametrized by $s$, which runs along the length of the DW, such that $\a(s)-\a(0)$ is an odd function of $s$ (i.e. the deformation has a reflection symmetry in a line along $\a(0)$). So in general, the energy difference is given by
\eq{
	\label{eq:ged}
	E_\textrm{el}=\int ds\mathcal{E}_\textrm{min}(\a(s))-L\mathcal{E}_\textrm{min}(0).
}
For a situation such as studied above, with $u/L\ll1$, we do a second order approximation of the right hand side in $u/L$. The result is
\eq{
	\label{eq:ged1}
	E_\textrm{el}\approx \frac{2u^2}{L}\left(\mathcal{E}_\textrm{min}(0)+\frac{\6^2\mathcal{E}_\textrm{min}}{\6\a^2}(0)\right).
}
On the right hand side of this equation the two elasticity components are made explicit. The first term is the stretching energy density and the second term the orientation energy density.

When considering bending, it becomes hard to determine $\mathcal{E}_\textrm{min}$ analytically. In the previous section we have computed the bending contribution from the exchange interaction to the energy density given by \cref{eq:eben}. To solve such a system one could expand the energy density around $\a_0=\a(0)$ and $\f_0=\f(0)$, which is the equilibrium angle corresponding to a DW orientation $\a(0)$ (so $\frac{\6\ve}{\6\f}(\a_0,\f_0)=0$):
\ea{
	\label{eq:ged2}
	\mathcal{E}(\a,\f)&\approx \mathcal{E}(\a_0,\f_0)+(\a-\a_0)\mathcal{E}_\a+\frac{1}{2}(\a-\a_0)^2\mathcal{E}_{\a\a}\\&+\frac{1}{2}(\f-\f_0)^2\mathcal{E}_{\f\f}+(\a-\a_0)(\f-\f_0)\mathcal{E}_{\a\f}, \nonumber
}
with $\mathcal{E}_\a=\frac{\6\mathcal{E}}{\6\a}(\a_0,\f_0)$ and $\mathcal{E}_{ab}=\frac{\6^2\mathcal{E}}{\6 a\6 b}(\a_0,\f_0)$.
If we neglect bending and solve the Euler Lagrange equation for $\f$, we find
\eq{
	\label{eq:fim1}
	\f=\f_0-(\a-\a_0)\frac{\mathcal{E}_{\a\f}}{\mathcal{E}_{\f\f}}.
}
Inserting this in \cref{eq:ged2} we get
\ea{
	\label{eq:ged3}
	\mathcal{E}(\a)\approx& \mathcal{E}(\a_0,\f_0)+(\a-\a_0)\mathcal{E}_\a+\frac{1}{2}(\a-\a_0)^2\mathcal{E}_{\a\a}\\
	&-\frac{1}{2}(\a-\a_0)^2\frac{\mathcal{E}_{\a\f}^2}{\mathcal{E}_{\f\f}}.\nonumber
}
To find the energy of the deformed wall we just have to integrate over $s$. As $\a(s)-\a(0)$ is odd, the second term vanishes upon integration. After subtracting the unperturbed DW energy we obtain the energy difference
\eq{
	\label{eq:ged4}
	E_\textrm{el}\approx \Delta L \ve(\a_0,\f_0)+S\left(\ve_{\a\a}-\frac{1}{2}\frac{\ve_{\a\f}^2}{\ve_{\f\f}}\right),
}
where $\Delta L=\int ds - L$ and $S=\int (\a(s)-\a_0)^2 ds$. Note that although this result is more general than the result of \cref{eq:ged2}, it is less accurate, because of the assumption that not only $\a(s)-\a_0$, but also $\f(s)-\f_0$ has to be small. 

When bending cannot be neglected the Euler Lagrange formalism gets an additional term which makes it a differential equation. 
\eq{
	\label{eq:fim2}
	\f-4\frac{J\l}{\mathcal{E}_{\f\f}}\frac{\6^2s}{\6\f^2}=\f_0-(\a-\a_0)\frac{\mathcal{E}_{\a\f}}{\mathcal{E}_{\f\f}}.
}
The solution depends on the profile $\a(s)$ of the DW. In \cite{Sok17} a circle segment shape is chosen and the equation is solved when there is no boundary condition for $\f$. At the end of this section we reproduce their model and extend it by optimizing for the segment length $L$ to find the free energy barrier. 

\begin{figure}[ht!]
	\center
	\includegraphics[width=\linewidth]{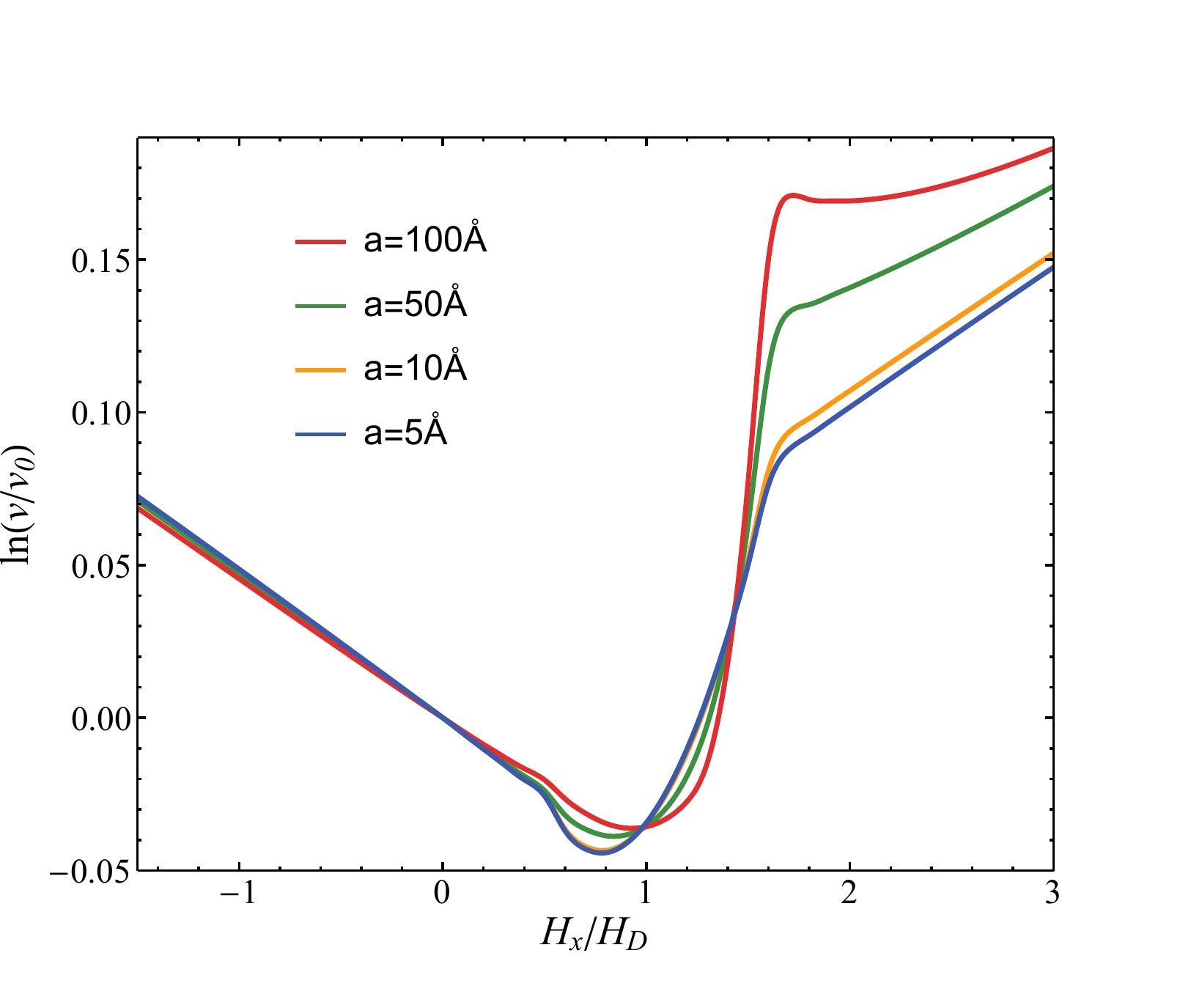}
	\caption{\label{fig:aef} The effective lattice spacing affects the DW dynamics.}
\end{figure}

However, as this solution does not account for bending at the boundaries, we first work out the model of this paper, which implements the bending energy in a rough and simplified manner: We choose the deformation to be triangularly shaped, determined by $u$ and $L$, following Blatter \textit{et al.}\cite{Bla94rev} The segments will have a constant value for $\f$ each and at the bending points a nearest neighbor exchange interaction is implemented. The elasticity is then enriched by the following bending energy contribution:
\eq{
	E_\textrm{bend}=\frac{J\l}{a}(3-\cos(\f_0-\f_1)-\cos(\f_0-\f_2)-\cos(\f_1-\f_2)).
}
Here $\f_0$ is the azimuthal angle of the magnetization of the unperturbed DW, and $\f_1$ and $\f_2$ are the optimized angles of the upper and lower segment respectively. The factor $3$ comes from the subtraction with the straight DW. Due to variations in the lattice structure and to account for non-nearest neighbor interactions, an effective value of $a$ should be used, but as demonstrated in \cref{fig:aef} the precise value of $a$ has no significant effect on the DW dynamics. The elasticity is now given by
\begin{widetext}
	\eq{
		\label{eq:eel2}
		\frac{E_\textrm{el}(u,L)}{t}=&\min_{\f_1,\f_2}\left[
		\frac{L}{2}\sqrt{1+\left(\frac{2u}{L}\right)^2}\left(
		\mathcal{E}(\a_0+\arctan(2u/L),\f_1)+\mathcal{E}(\a_0-\arctan(2u/L),\f_2)
		\right)
		\right.\\
		&\left.+\frac{J\l}{a}(3-\cos(\f_0-\f_1)-\cos(\f_0-\f_2)-\cos(\f_1-\f_2))
		\right]-L\mathcal{E}(\a_0,\f_0).
	}
After minimizing the elasticity in \cref{eq:eel2} over $\f_1$ and $\f_2$, one obtains the elasticity as a function of $u$ and $L$. This dispersive elasticity is used in the main article as a starting point for our model. The energy density is given by \cref{eq:endendl} and is made dimensionless by dividing out $D$:
	\eq{
		\label{eq:endendl}
		\tilde{\mathcal{E}}(\a,\f)\equiv\frac{\mathcal{E}(\a,\f)}{D}=2\tilde{J}\sqrt{1-\b^2}+\p \left(g(\b)\frac{\tilde{H_\textrm{B}}}{2}\cos^2(\f-\a)-f(\b)\tilde{H_x}\cos(\f)+f(\b)\cos(\f-\a)\right),
	}
with $\tilde{J}\equiv J \lambda^{-1} D^{-1}$, $\tilde{H_\textrm{B}}\equiv2H_\textrm{B}/H_\textrm{D}$ and $\tilde{H_x}\equiv H_x/H_\textrm{D}$.
By assuming $u\ll L$, setting $\f_1=\f_2=\f_0$ and $\a_0=0$, the elasticity is approximated to be proportional to $u^2/L$: $E_\textrm{el}\approx\frac{u^2}{L}\mathcal{E}_{\textrm{el},0}$ with
\eq{
\mathcal{E}_{\textrm{el},0}=4\tilde{J}\sqrt{1-\b^2}+
\begin{cases}
	+2\p f(\b)|\tilde{H_x}|-\p g(\b)\tilde{H_\textrm{B}} \qquad\text{, if }|\tilde{H_x}-1|>\tilde{H_\textrm{B}};\\
	-\p\frac{f(\b)^2}{\tilde{H_\textrm{B}}g(\b)}(\tilde{H_x}-1)(\tilde{H_x}-3) \qquad\text{, otherwise.}
\end{cases}
}
\end{widetext}
This formula is used to fit data using $\ln(v)\propto-(\mathcal{E}_{\textrm{el},0}/H_z)^{1/4}$. However, due to the severe assumptions made to obtain it, the fit will not be good and should be used to obtain good guesses for the relevant parameters. To determine the exact velocity we continue with the exact elasticity of \cref{eq:eel2}. The optimization over $\f_1$ and $\f_2$ is done numerically.

\begin{figure}[ht!]
	\center
	\includegraphics[width=\linewidth]{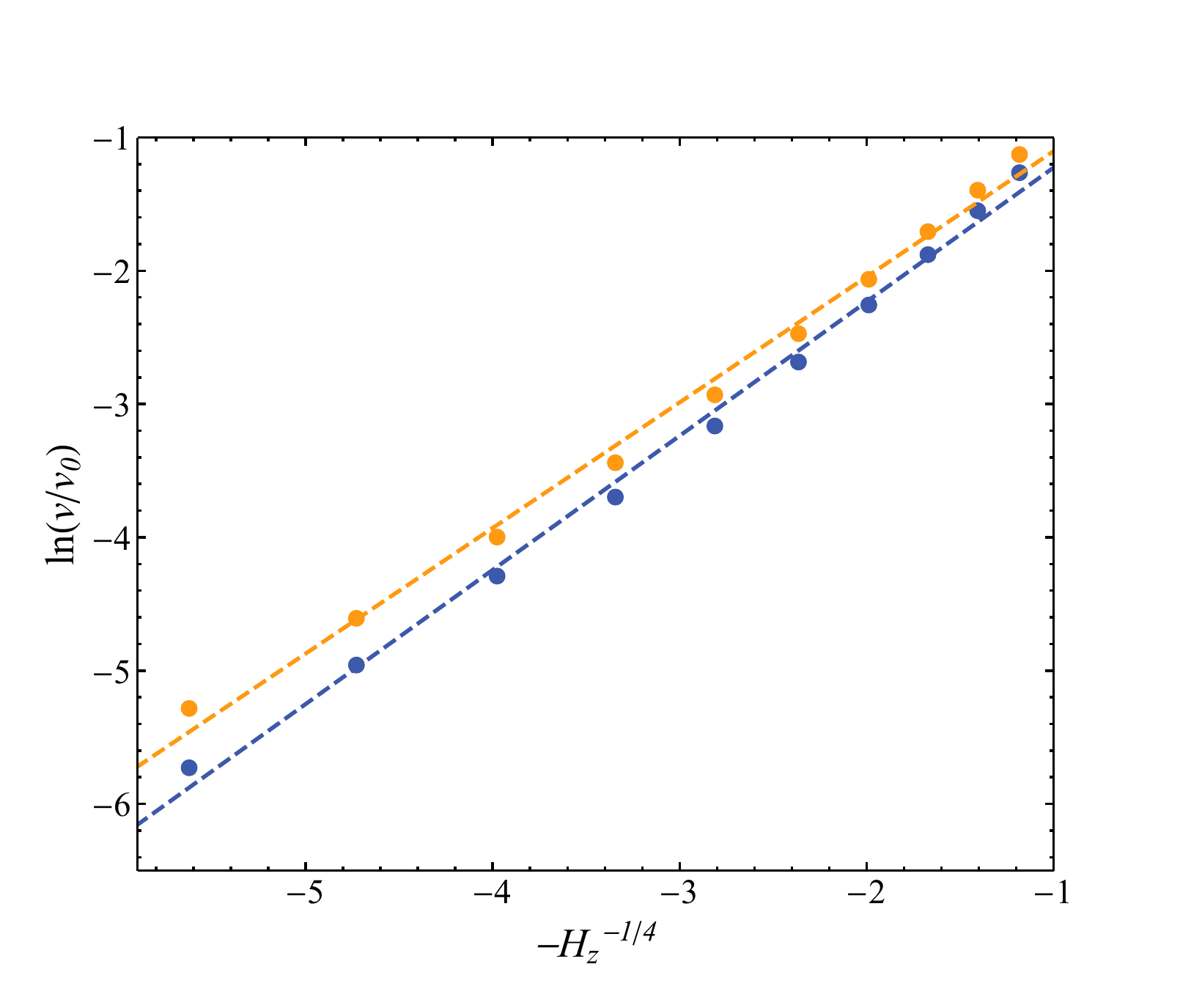}
	\caption{\label{fig:hzc} The exact creep velocity as a function $H_z^{-1/4}$ for $\tilde{H_\textrm{B}}=1.3$, $\tilde{H_x}=0.5$ (blue) and $\tilde{H_x}=2$ (orange). Even the exact model deviates from the famous creep law.}
\end{figure}

The Larkin length now is found as usual inserting $u=\xi$ as described above. However, this has to be done numerically as well as the computation of the free energy barrier. The result is shown and discussed in the main paper. In \cref{fig:hzc} we plot $\ln(v)$ as a function of $H_z^{-1/4}$ which, in our exact theoretical model, is not perfectly linear.

Instead of considering a DW normal to the applied IP magnetic field, one could tilt the DW itself over a fixed angle $\a$ as suggested by Pellegren et. al. \cite{Sok17}. From \cref{eq:endendl} it becomes clear that the pure Néel DW then no longer minimizes the energy density for $|H_x-H_\textrm{D}|>2H_\textrm{B}$; the Néel DW is the asymptotic limit as $H_x\rightarrow\infty$. Because there will be no saturation of the azimuthal angle $\f$, the velocity profile becomes more smooth. As described in the main text, we do not need to interpret a fixed value of $\a$ as a physical tilting of the DW, but use it solely to account for higher order anisotropy terms.


\begin{figure}[ht!]
	\center
	\includegraphics[width=4cm]{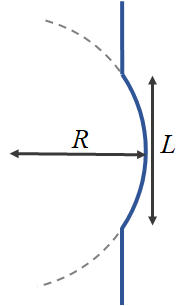}
	\caption{\label{fig:pelsys} An illustration of the deformation model considered by Pellegren \textit{et al.}\cite{Sok17}}
\end{figure}
To conclude this section, we reproduce and expand on the results from Pellegren \textit{et al.}, who considered an arc shape deformation parametrized not by the displacement $u$, but by the radius of the circle segment illustrated in \cref{fig:pelsys}. The relation between $u$ and $R$ is given by
\eq{
	\label{eq:Ru}
R=\frac{L^2+4u^2}{8u}.
}
Using \cref{eq:ged2} enriched with \cref{eq:eben}, we find $\f$ by the Euler Lagrange formalism. The boundary conditions are not given by a restriction on the value of $\f$ at the boundary of the segment, but by minimizing the energy. Assuming $R\ll L$ the solution for $\f$ cen be approximated by:
\eq{
\f(s)=\f_0+\frac{\mathcal{E}_{\a\f}}{\mathcal{E}_{\f\f}}\frac{1}{R}\left(s-\Lambda \sinh(s/\Lambda)\right).
}

Now we compute the elasticity as a function of $R$ and $L$. Using the Larkin length and the wandering exponential relation \cref{eq:wex} combined with \cref{eq:Ru}, the elasticity is only a function of $L$. Then the optimal length $L_\textrm{opt}$ is computed as a function of $H_x$ as well as the corresponding free energy barrier. 

\begin{figure}[t]
	\center
	\includegraphics[width=\linewidth]{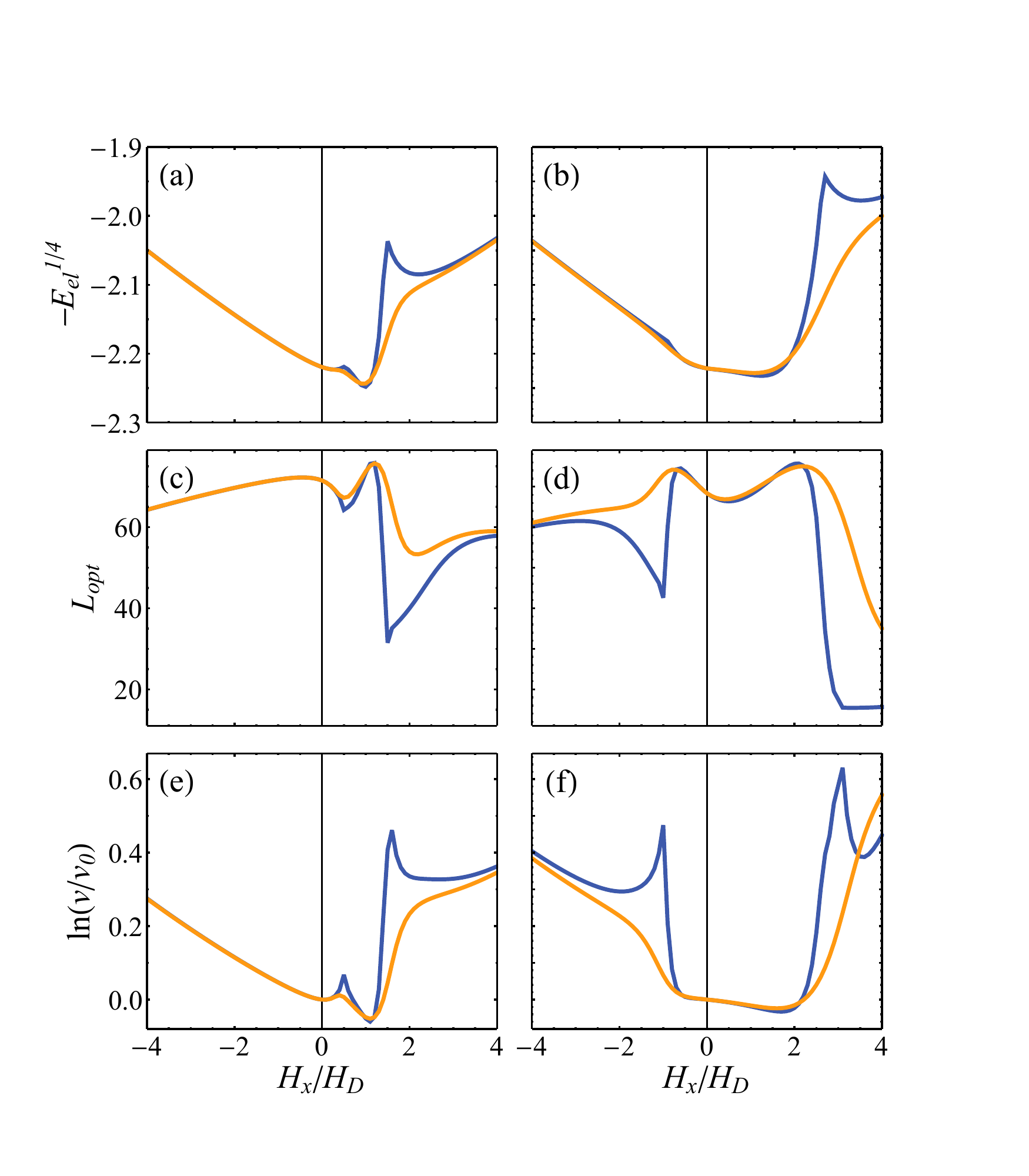}
	\caption{\label{fig:sok} The exact results for the arc deformation model proposed by Pellegren \textit{et al.} \cite{Sok17} for $\a_0=0$ (blue) and $\a_0=8^\circ$ (orange). The two columns correspond to $\tilde{H_\textrm{B}}=0.5$ (left) and $\tilde{H_\textrm{B}}=2$ (right) . The optimized deformation length $L_\textrm{opt}$ is not constant as a function of $H_x$ (a) and (b), as a result the velocity profiles (c) and (d) differ form the elasticity with $L(H_x)=L$ fixed shown in (e) and (f) respectively. Furthermore, due to the saturation of $\f_0$ at $\a_0=0$, the results again show kinks.}
\end{figure}

In \cref{fig:sok} this optimal length is plotted along with the DW velocity computed from the corresponding energy barrier. In the same figure the elasticity for a fixed value of $L$ is also shown to emphasize the effect of optimizing for $L$. These results are generated for $\a_0=0$ (left column) and $\a_0=8^\circ$ (right column). Notice that for $\a_0=0$ the same kinks occur, which has the same cause: the saturation of $\f_0$.

\clearpage
\section{Experimental method}
\label{app:F}
The samples are grown via Ar DC magnetron sputter deposition in a sputter chamber with a base pressure of  $\sim3\times10^{-9}$ mbar. The detailed composition of the samples introduced in the main article is:

\begin{tabular}{l|l}
Sample a
&
SiO$_2$/Ta(4)/Pt(4)/Co(0.6)/Pt(4)
\\
\hline
Sample b
&
SiO$_2$/Ta(4)/Pt(4)/Co(0.8)/Gd(4)
\\
\hline
Sample c
&
SiO$_2$/Ta(4)/Pt(4)/Co(0.9)/Ir(4)
\end{tabular}
The number in parentheses indicates the thickness of the layer in nanometers. These samples are representatives of the variety of the velocity profiles observed in the literature of asymmetric domain expansion experiments.\cite{Je13,Lav15,Van15,Lau16}.

The data of sample (a) was obtained from Ref.~[\onlinecite{Lav15}]. The sample is a symmetric stack where in an ideal case no DMI should be present. A finite DMI is obtained when the difference of the interfacial quality of the sputtered layers is considered. The authors studied the interfacial quality by changing the growth pressure of the top platinum layer. A growth pressure of $1.12$ Pa is used for the top layer in sample (a). 

The data of samples (b) and (c) are part of a larger sample study performed for this paper on the effect of the Cobalt layer thickness. The results of the sample study are shown in the following section.

For all samples a perpendicular magnetic anisotropy is determined via angle-dependent anomalous Hall effect measurements. The saturation magnetization $M_\textrm{S}$ is obtained via SQUID-VSM measurements. We image the magnetic domains and the expansion of those domains with a Kerr microscope setup. In this setup a difference in contrast in the image corresponds to a different magnetization in the z-direction. A dark contrast in the images indicates that the magnetization is pointing into the plane, referred to as a down domain. A light contrast in the image corresponds to an up domain. These domains are nucleated around intrinsic impurities and defects in the sample, by applying out-of-plane magnetic field pulses. The pulses can vary in length between $0.8-400$ ms and can reach a field strength up to $\pm33$ mT. 
\begin{figure}
	\includegraphics[width=\linewidth]{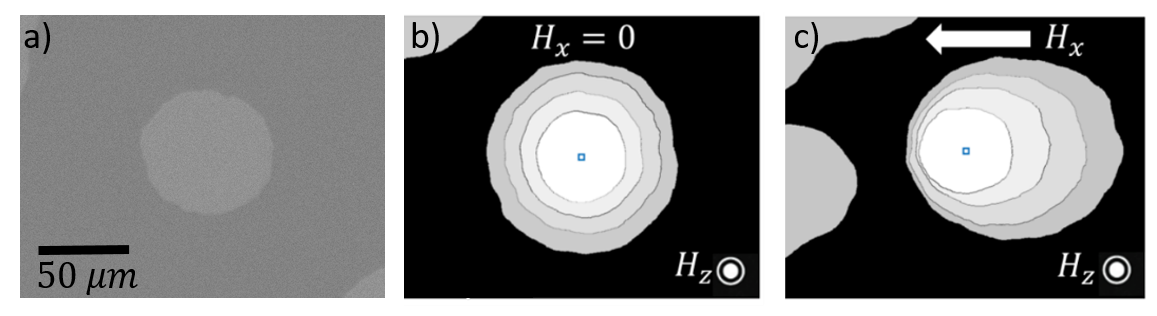}
	\caption{\label{fig:met} Kerr microscope images of a magnetic domain. The magnetization in the light areas is pointing out of the plane (up direction) and the magnetization in the black areas is pointing into the plane (down direction). In (a) an unprocessed Kerr image is shown. A binary image is obtained via thresholding. In (b) the nucleation of the domain is shown in white and the three expansion steps with a darkening gray scale. Only out-of-plane magnetic field pulses are applied and this results in a symmetric expansion of the magnetic domain to all sides. In (c) an additional in-plane field is applied in the x-direction and this results in an asymmetric expansion along that direction.}
\end{figure}

In \cref{fig:met} (a) the nucleation of an up domain is shown. The image is processed to a binary image by thresholding. After the nucleation the domain is expanded in three steps via additional out-of-plane field pulses. In figure \cref{fig:met} (b) the nucleation of an up domain is shown in white and the three expansion steps are indicated by a darkening gray scale. We find that the expansion of the magnetic domain is symmetric to all sides. In \cref{fig:met} (c) an in-plane magnetic field along the $x$-direction is added during the expansion of the domain. In this case, the right side of the domain expands much faster than the left side, resulting in an asymmetric magnetic domain expansion. The velocity of the domain boundary is obtained by measuring its displacement for a known pulse length. Repeating this as a function of the in-plane field $H_x$ results in a velocity profile.  

\section{Sample study: Effect of ferromagnetic layer}
\begin{figure}[b]
	\includegraphics[width=\linewidth]{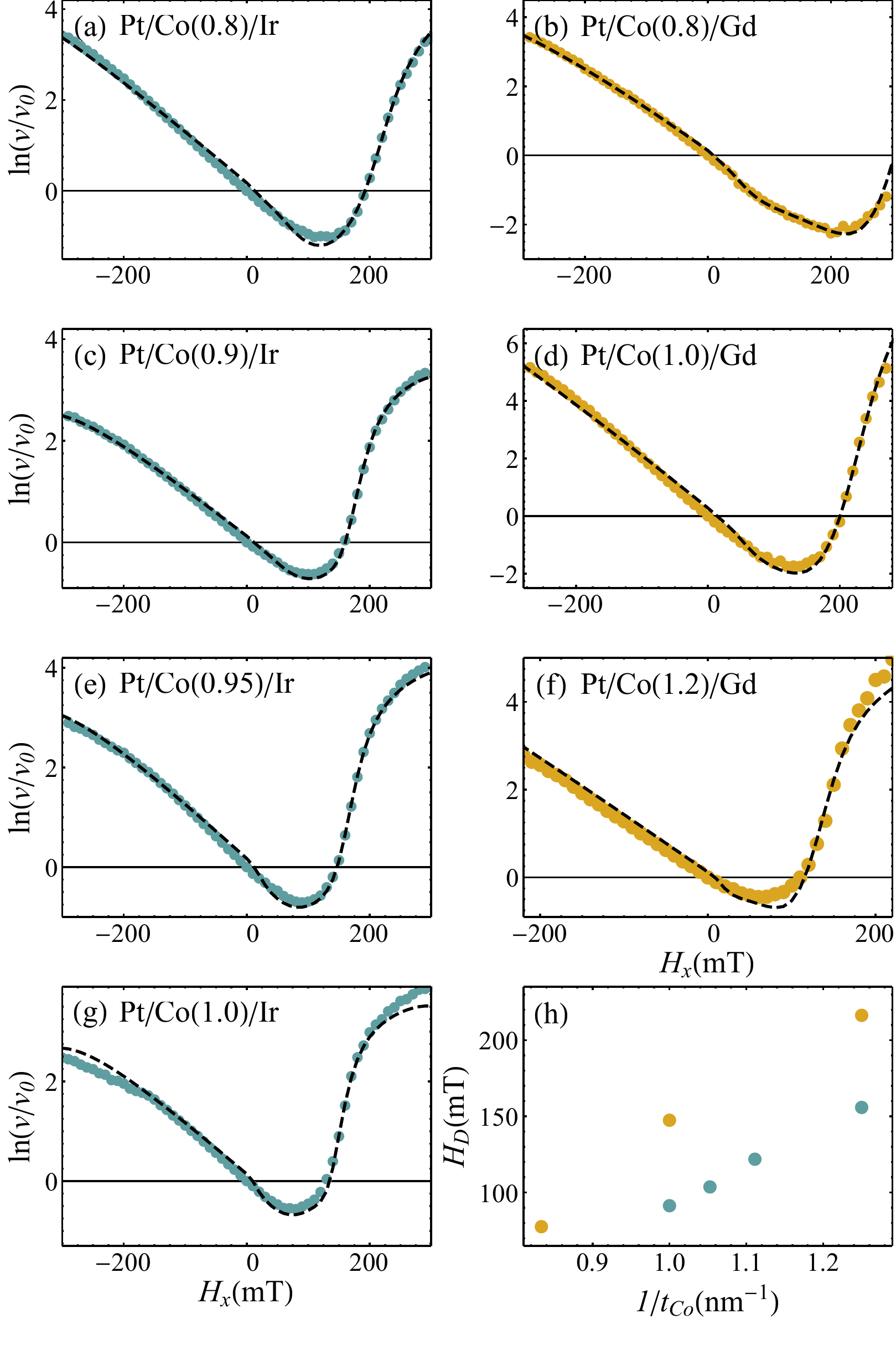}
	\caption{\label{fig:HDt} Effect of ferromagnetic layer thickness on the velocity profile as a function of $H_x$. The data (dots) are fitted (dashed lines) with our model to obtain a value for the effective DMI field $H_\textrm{D}$, summarized in (h). (a), (c), (e) and (g) show the results for the Pt/Co(x)/Ir stacks and (b), (d) and (f) show the results for the Pt/Co(x)/Gd stacks with varying thickness of the Co layer.}
\end{figure}

We applied our model to determine the effect of the layer thickness of the ferromagnet on the strength of the DMI. Since the DMI is an interfacial effect, we would expect that the DMI decreases for an increase in the cobalt layer thickness. We studied two types of stacks: SiO$_2$/Ta(4)/Pt(4)/Co(x)/Gd(4) and SiO$_2$/Ta(4)/Pt(4)/Co(x)/Ir(4). The result is shown in \cref{fig:HDt}.

The obtained trends are in line with our expectations described in the first section of this supplemental material (i.e. that $H_\textrm{D}\propto t^{-1}$) and compare well to previous experiments. \cite{Thi12,Bel18} Future research could focus on a larger sample study with a thorough investigation of uncertainties in the measurement and determination of the sample parameters (such as the thickness of the saturation magnetization) to establish a proportionality $D\propto t^{-1}$.
\clearpage
\begin{widetext}
\section{Constants and parameters}
\label{app:E}
\begin{table}[h!]
\begin{tabular}{l|p{12cm}|r}
	Constant & Description & Value\\
	\hline
	$\a$ & angle of the IP component of the DW normal w.r.t. the IP magnetic field & - \\
	$\b$ & parametrization of the tilted magnetization inside mangetic domains: $\sin(\th)=\b$ & -\\
	$\Delta$ & pinning energy constant & $10^3$J$^2/$m$^3$\\
	$\z$ & wandering exponent: $\la\la (u(y+L)- u(y))^2\ra\ra\propto(L/L_\textrm{c})^{2\z}$ & 2/3 \\
	$\th$ & angle of the internal magnetization of the DW with respect to the $z$-axis & -\\
	$\lambda$ & DW width & $5$nm\\	
	$\xi$ & correlation length of the pinning potential & $10$nm\\
	$\phi$ & IP angle of the internal magnetization of the DW with respect to the $x$-axis & -\\
	$A$ & assymetric component of the DW velocity: $A=\ln(v(\uparrow\downarrow)/v(\downarrow\uparrow))$ & -\\
	$a$ & (effective) lattice spacing & $2$nm\\
	$b$ & defined from $b=\arccos(\b)/\p$ & -\\
	$D$ & DMI energy defined from $D=D_\textrm{s} a^{-1}t^{-1}$ & -\\
	$\mathcal{E}$ & DW energy density & -\\
	$\mathcal{E}_\textrm{el}$ & the elasticity of the DW (energy cost per length) & - \\
	$F_\textrm{b}$ & free energy barrier associated with a DW deformation & -\\
	$H_\textrm{B}$ & effective field favoring the Bloch wall arising from dipole-dipole interaction & -\\
	$H_\textrm{D}$ & effective DMI field & -\\
	$H_\textrm{DW}$ & effective field defined as the $\f$ dependent part of $\mathcal{E}$ & -\\
	$H_x$ & externally applied IP magnetic field along the $x$-axis & -\\
	$H_y$ & externally applied IP magnetic field along the $y$-axis & -\\
	$H_z$ & externally applied OOP magnetic field along the $z$-axis driving the DW & $10$mT \\
	$J$ & exchange energy defined from the spin exchange $J=J_\textrm{s}/a$ & $2\times 10^{-11}$J/m \\
	$K_\textrm{P}$ & PMA energy defined as $K_\textrm{P}=J/\l^2$ & $8\times 10^5$ J/m$^3$\\
	$L$ & length of a thermally fluctuating DW segment & -\\
	$M_\textrm{S}$ & saturation magnetization & $5\times 10^5$ A/m \\
	$T$ & Temperature & $300 ^\circ$K\\
	$t$ & film thickness & $0.5$nm\\
	$u$ & displacement of a thermally fluctuated DW segment normal to the DW & -\\
	$u_\textrm{c}$ & proportionality constant of the wandering relation & $10$nm
\end{tabular}
\caption{\label{tab:cnp}Desciption of the constants and symbols used in this paper and their typical values if applicable.}
\end{table}
\end{widetext}
	\bibliography{References}